\documentclass[11pt,preprint]{aastex}
\newcommand{\bl}[1]{\mbox{\boldmath$ #1 $}}
\newcommand{\cs}{c_{\rm s}}
\newcommand{\csq}{c^2_{\rm s}}
\newcommand{\zhat}{\mbox{\boldmath$ \hat{z}$}}
\newcommand{\rhat}{\mbox{\boldmath$ \hat{r}$}}
\newcommand{\phihat}{\mbox{\boldmath$ \hat{\phi}$}}
\newcommand{\beq}{\begin{equation}}
\newcommand{\eeq}{\end{equation}}

\newcommand{\Msun}{M_{\odot}}
\newcommand{\ul}{\underline{\hspace{20pt}}}

\slugcomment{to appear in ApJ 20 Oct 2006}
\shorttitle{Burst Mode of Accretion}
\shortauthors{E. I. Vorobyov and Shantanu Basu}

\begin{document}

\title{The Burst Mode of Protostellar Accretion}

\author{E. I. Vorobyov\altaffilmark{1,}\altaffilmark{2}
and Shantanu Basu\altaffilmark{2}}
\altaffiltext{1}{CITA National Fellow.}
\altaffiltext{2}{Department of Physics and Astronomy, University of 
Western Ontario, London, Ontario,  N6A 3K7, Canada; 
vorobyov@astro.uwo.ca, basu@astro.uwo.ca.}

\begin{abstract}
We present new numerical simulations in the thin-disk approximation 
which characterize the burst mode of 
protostellar accretion. The burst mode begins upon the formation
of a centrifugally balanced disk around a newly formed protostar.
It is comprised of prolonged quiescent periods of low 
accretion rate (typically $\la 10^{-7} \Msun$ yr$^{-1}$) which are
punctuated by intense bursts of accretion (typically $\ga 10^{-4} \Msun$ yr$^{-1}$,
with duration $\la 100$ yr)
during which most of the protostellar mass is accumulated.
The accretion bursts are associated with the formation of dense
protostellar/protoplanetary
embryos, which are later driven onto the protostar by
the gravitational torques that develop in the disk.
Gravitational instability in the disk, driven by continuing infall from 
the envelope, is shown to be an effective means of 
transporting angular momentum outward, and mass inward to the protostar.
We show that the disk mass always remains significantly less than
the central protostar mass throughout this process.
The burst phenomenon is robust enough to occur
for a variety of initial values of rotation rate, frozen-in (supercritical)
magnetic field, and density-temperature relations. Even in cases 
where the bursts are nearly entirely suppressed, a moderate increase in
cloud size or rotation rate can lead to vigorous burst activity.
We conclude that most (if not all) protostars undergo a burst mode
of evolution during their early accretion history,
as inferred empirically from observations of FU Orionis variables.

\end{abstract}

\section{Introduction}
Many stars in the pre-main sequence evolutionary stage are known to vary 
in brightness. 
For instance, T~Tauri and Herbig Ae/Be stars change in brightness by about 
one magnitude in irregular intervals and EXors show short-lasting 
variations of 1-3 magnitudes.
However, the stars with the largest
amplitude brightness variations (3-6 magnitudes during $\le 100$~yr) 
are the FU Ori stars named after 
the prototype FU Orionis \citep{Herbig}.
A sharp increase in the mass accretion rate onto a protostar is widely
thought to be responsible for the FU Ori outbursts. 

Several possible physical mechanisms for elevated
mass accretion rates in young stars have been proposed.
For instance, a strong dependence of the effective viscosity on the gas
temperature in the innermost regions of a circumstellar disk can cause a
temporary increase in the viscous mass transfer rate and generate mass 
accretion bursts \citep*{Lin,Clarke}.
The thermal instability theory has been refined to successfully
model the duration and inferred repetition timescale of FU Ori
outbursts, and even to model individual light curves of a few objects
\citep{Bell94,Bell95}.
Nevertheless, these models depend upon the ad hoc $\alpha$ viscosity
prescription \citep{SS}, which parametrizes the physical mechanism
of angular momentum transport.
Another possible cause of FU Ori outbursts is a close encounter
of protostars in a binary system, resulting in tidal effects in the disks
and mass transfer \citep{Bonnell}. This mechanism obviously fails to
explain the FU Ori phenomenon in isolated protostars.

In recent years, a new mechanism for the generation of FU Ori eruptions has
begun to emerge. Theoretical and numerical
studies of protostellar disks show that they are subject
to gravitational instabilities which redistribute
the mass and angular momentum in the disk \citep{Tomley,Laughlin}.
\citet{Boss2} has found that the gravitational instability in marginally
unstable disks can lead not only to a steady inward mass transport but also to 
protoplanet formation. 
\citet{Mejia} have reported
three-dimensional disk simulations that demonstrate
a single FU-Ori-like mass accretion burst related to the growth of
spiral instability. The above-mentioned models all study the evolution of an 
{\it isolated} protostellar or protoplanetary disk. 
In our recent paper \citep[hereafter Paper I]{VB3}
we presented the first model of cloud core collapse which
generates {\it episodic} mass accretion and luminosity bursts 
that can be identified with FU Ori eruptions.  The disk that forms 
self-consistently in our model from the collapse of a cloud core is
necessarily {\it not isolated} from its parent cloud core envelope.
The results presented in Paper I showed that
the infall of matter from the surrounding envelope 
can periodically destabilize the disk and lead to the formation of
spiral structure and dense clumps within the
arms. Gravitational torques associated with the spiral arms drive the 
clumps onto the protostar, generating  mass accretion amd luminosity bursts 
comparable to those observed in FU Ori stars. This process repeats until most 
of the envelope has been accreted onto the protostellar disk.
We note that this process bears a remarkable resemblance to the empirically
inferred history of disk accretion advocated by \cite{Kenyon}, and 
reviewed by e.g., \citet{Hartmann2}.

We refer to the newly discovered mode of protostellar accretion as the
``burst mode''. It follows an earlier ``smooth mode'' of accretion that
is characteristic of mass accretion onto the protostar before a disk
has formed. The smooth mode accretion rate may continue to be used 
as a value for envelope accretion onto the disk after its formation.
The smooth mode accretion has been extensively studied 
in the literature, accounting for the influence of various 
initial conditions and physical processes. 
In the prototypical model of 
smooth mode accretion \citep{Shu77}, the accretion rate is proportional 
to $\cs^3/G$, where $\cs$ is the isothermal sound speed
and $G$ is the gravitational constant. The proportionality constant is
0.975 in the case of a static equilibrium singular isothermal sphere (SIS)
density profile ($\rho = \cs^2/2 \pi G r^2$) at the moment of protostar 
formation ($t=0$). However, it is increasingly greater for profiles that are overdense
compared to the SIS but still proportional to $r^{-2}$. The effect of
nonzero but spatially uniform velocities at $t=0$ is to increase the
accretion rate further \citep{Hunter,Fatuzzo}, while remaining
proportional to $\cs^3/G$. The effect of spatially nonuniform velocities
that are commonly generated in models of prestellar collapse
is to introduce a time dependence into the accretion rate. There is an
initially high accretion rate that is reminiscent of the 
self-similar solution derived
by \citet{Hunter} -- due to an inner region of supersonic infall at $t=0$ --
but a subsequent decline and later accretion rate 
that is closer to the Shu rate \citep{Hunter,FC}. 
The effect of magnetic fields is to increase
the constant accretion rate somewhat in the case of static initial conditions
\citep{LS} and to increase the time-dependent accretion rate for 
nonzero and spatially nonuniform infall at $t=0$
\citep{Tomisaka,CK98}.
A finite mass reservoir will clearly introduce a final phase of 
terminally declining accretion rate. Therefore, the smooth mode accretion can
be conceptually divided into three phases \citep{VB1,VB2}:
an early phase of high but declining accretion rate,
an intermediate phase of steady accretion at a rate proportional to 
$\cs^3/G$ (this phase may not exist if the cloud core is sufficiently small),
and a final phase of terminally declining accretion rate.

In the present paper, we investigate further the effect of rotation and 
nonaxisymmetry in generating the burst mode of accretion. For better study of
the influence of rotation and comparison with spherically symmetric models, 
we also present nonrotating models.
Furthermore, we investigate the role of frozen-in magnetic fields and thermal evolution
in determining the frequency and amplitude of mass accretion bursts.
We show how the burst mode of protostellar accretion 
is related to the three phases of smooth mass accretion,
since the latter still characterize
the infall onto a protostellar disk after its formation.

The basic equations that we solve and the numerical techniques are
described in \S~\ref{formulation}, while the results of our simulations
are presented in \S~\ref{results}. Further discussion is 
given in \S~\ref{discussion} and the results are summarized in
\S~\ref{summary}. The Appendix describes some tests of our numerical code.

\section{Formulation of the Problem}
\label{formulation}

\subsection{Basic Equations}
\label{model}
We consider cloud cores that have dynamically decoupled from a diffuse 
parent cloud,
and which are flattened due to the influence of an energetically significant
large-scale magnetic field and/or rotation. 
We use the magnetized thin-disk approximation \citep{Ciolek,Basu94} 
generalized to nonaxisymmetric form.
The thin-disk approximation is justified 
certainly during the prestellar collapse phase, as the core 
quickly establishes near force-balance along the mean magnetic field direction
\citep{Fiedler}.  
We also assume a spatially uniform 
mass-to-flux ratio, which is approximately valid for much of the 
supercritical phase of collapse \citep{Basu}. 

The basic model is as follows. In the vertical ($z$) direction coinciding with the 
mean magnetic field, we use a one-zone approximation, so that the cloud of half thickness
$Z(r,\phi,t)$ contains a vertically uniform density $\rho(r,\phi,t)$ and vertically uniform 
magnetic field $B_z(r,\phi,t) \zhat$, where $r$ and $\phi$ are the radial and
azimuthal coordinates in the disk equatorial plane, and $t$ is the time. 
The magnetic field has nonzero components in all
three directions above the disk surface. In particular, at the top ($z=Z$) 
and bottom ($z=-Z$) surfaces of the cloud, the magnetic field is
\begin{eqnarray}
\bl{B}(r,\phi,z=Z,t) & = & B_z^+\zhat + B_r^+\rhat + B_{\phi}^+\phihat, \\
\bl{B}(r,\phi,z=-Z,t) & = & B_z^-\zhat + B_r^-\rhat + B_{\phi}^-\phihat .
\end{eqnarray}
The physical variables on the right hand side of the above equations are functions
of $(r,\phi,t)$; all thin-disk variables subsequently listed will be assumed to carry 
this dependence.
To simplify further, we note that reflection symmetry about the midplane implies that
$B_z^- = B_z^+$, $B_r^- = -B_r^+$, and $B_{\phi}^- = -B_{\phi}^+$. 
We shall also assume that the vertical component of the magnetic field inside the disk
equals that just outside, i.e., $B_z^+ = B_z$; this actually ignores a correction
term proportional to $\nabla_p Z$ \citep{Ciolek}, 
where $\nabla_p = 
\rhat \partial/\partial r + \phihat \, r^{-1}  \partial/\partial \phi$ is the
gradient along the planar coordinates of the disk. Terms that are proportional
to $\nabla_p Z \sim Z/r$ amount to small corrections to the thin disk equations, 
and are generally dropped for simplicity in this study. 
For ease of notation, we subsequently drop all $+$ 
superscripts for magnetic field components at the disk surface.

Integration of the ideal MHD equations in the $z-$direction, from $z=-Z(r,\phi,t)$ to $z=Z
(r,\phi,t)$ and use of Leibnitz's Rule for Differentiation of Integrals and the
Fundamental Theorem of Calculus, yields the following equations for mass and momentum
transport:
\begin{eqnarray}
\label{cont}
 \frac{{\partial \Sigma }}{{\partial t}} & = & - \nabla _p  \cdot \left( \Sigma \bl{v}_p \right), \\ 
\label{mom}
 \Sigma \frac{d \bl{v}_p }{d t}  & = &  - \nabla _p {\cal P} - \frac{Z}{{4\pi }}\nabla _p B_z^2  + \frac{{B_z \bl{B}_p }}{{2\pi }} + \Sigma \bl{g}_p \, . 
\end{eqnarray}
In the above equations, $\Sigma$ is the mass surface density of the disk, 
${\cal P} \equiv \int_{-Z}^{Z} P dz$ is the vertically integrated 
form of the gas pressure $P$,
$\bl{v}_p = v_r \rhat + v_{\phi} \phihat$ is the velocity in the 
disk plane, $\bl{B}_p = B_r \rhat + B_{\phi} \phihat$ is the planar magnetic
field at the top surface of the disk, and $\bl{g}_p = g_r \rhat + g_{\phi}\phihat$ is
the gravitational field in the disk plane. Both $\bl{v}_p$ and $\bl{g}_p$ are assumed to be
$z-$independent for simplicity. 
In this derivation, we have 
neglected the effect of an external bounding pressure on the cloud, and dropped all
terms proportional to $\nabla_p Z$ in the integrated magnetic force.
Note the formal similarity between the magnetic pressure ($-2 Z \nabla_p B_z^2/8 \pi$) and 
thermal pressure ($- \nabla_p {\cal P} \simeq -2 Z \nabla_p P$) terms, 
as well as between the magnetic tension
($B_z \bl{B}_p/2 \pi$) and gravity ($-g_z \bl{g}_p/2 \pi G$)
terms for a thin sheet, in which $g_z = - 2 \pi G \Sigma$.

In order to calculate the forces due to gas pressure and magnetic pressure, 
we have to specify the half-thickness $Z$ in the effectively optically thin
(isothermal) and optically thick (nonisothermal) regimes. 
During the isothermal phase of evolution,
local vertical hydrostatic equilibrium (ignoring external gas and 
magnetic field pressure) implies that 
\beq
\label{forcebal}
\rho c^2_{\rm s} = \frac{\pi}{2} G \Sigma^2, 
\eeq
where $\cs$ is the isothermal sound speed. This relation allows
a straightforward determination of the 
half-thickness $Z \equiv \Sigma/2\rho$.
For the optically thick regime, we make use of
the radiation hydrodynamic simulations of spherical cloud collapse by
\citet{Masunaga98}, who have shown that the size of the central
optically thick region stays constant at approximately 5~AU and is
independent of the mass of a parent cloud and of the initial density distribution.
Therefore, we seek to capture the essential physics of both regimes by letting
\beq
Z = \left\{
\begin{array}{ll}
\csq/\pi G \Sigma & {\rm if}  \:\:  \Sigma < \Sigma_{\rm cr}, \\
\csq/\pi G \Sigma_{\rm cr} & {\rm if}  \:\: \Sigma \geq \Sigma_{\rm cr}.
\label{halfthick}
\end{array}
\right.
\eeq
We set the critical surface density $\Sigma_{\rm cr}=36.2$~g~cm$^{-2}$, which 
corresponds to the critical gas volume density 
$n_{\rm cr}=10^{11}$~cm$^{-3}$ \citep{Larson}
for a gas disk in vertical hydrostatic equilibrium at $T=10$~K. 
The dependence of our
results on the adopted value of $\Sigma_{\rm cr}$ is discussed in 
\S~\ref{freeparam}.

We assume isothermal evolution up to some critical density, and a 
polytropic pressure-density relation for the later optically thick regime.
This assumption, along with the adopted half-thickness $Z$ of the thin-disk, 
can be used to derive the following simplified expression for the integrated gas 
pressure as a function of surface density: 
\begin{equation}
{\cal P}=c_{\rm s}^2\Sigma
+ c_{\rm s}^2 \Sigma_{\rm cr} \left(\Sigma \over \Sigma_{\rm cr} \right)^\gamma.
\label{eos}
\end{equation}
Equation (\ref{eos}) allows for a smooth 
transition between the isothermal and nonisothermal regimes.  
The gas temperature can be determined via the ideal gas equation of state 
${\cal P}=\Sigma k T/m$, so that 
\begin{equation}
T={c^2_{\rm s} m \over k}\left[1  + \left( \Sigma \over \Sigma_{\rm cr} \right)^{\gamma-1} \right],
\label{tempr}
\end{equation}
where $k$ is Boltzmann's constant and $m$ is the mean molecular mass.
We adopt the ratio of specific heats $\gamma=7/5$ for the optically
thick regime, appropriate for an adiabatic diatomic gas.
If the temperature is low, neither 
rotational nor vibrational modes of molecular
hydrogen are excited. In this case, the ratio of specific heats is 
$\gamma=5/3$. We explore the dependence of our results on the
adopted value of $\gamma$ in \S~\ref{freeparam}.

The gravitational field $\bl{g}(r,\phi,z,t) = - \nabla \Phi(r,\phi,z,t)$, where
$\Phi$ is the scalar gravitational potential. 
If the material outside the thin disk is assumed to contain negligible
mass, then $\Phi$ is formally the solution of Laplace's equation
\beq
\nabla^2 \Phi =0 
\label{laplacegrav}
\eeq
above the disk.
For this purpose, we work in the limit
$Z \rightarrow 0$, in which case
the solution is subject to the boundary conditions
\beq
\frac{\partial \Phi}{\partial z} =  2 \pi G \Sigma \: (z = 0), \: \: 
\Phi \rightarrow 0 \: (z \rightarrow \infty). 
\label{gravbc}
\eeq
The gravitational field in the plane of the thin sheet is
\beq
\bl{g}_p(r,\phi,t) = - \nabla_p \Phi(r,\phi,z=0,t).
\label{gp}
\eeq
The solution for the 
potential in the plane of the nonaxisymmetric sheet is 
\begin{eqnarray}
  \Phi(r,\phi,z=0,t) & = & - G \int_0^{r_{\rm out}} r^\prime dr^\prime 
                      \nonumber \\
      & &        
                \times  \int_0^{2\pi} 
               \frac{\Sigma(r^\prime,\phi^\prime) d\phi^\prime}
                    {\sqrt{{r^\prime}^2 + r^2 - 2 r r^\prime
                       \cos(\phi^\prime - \phi) }}  \, ,
\label{potential}
\end{eqnarray}
where $r_{\rm out}$ is the size of the cloud core 
\citep[see][]{BT}.

Due to our assumption of flux-freezing and a spatially uniform mass-to-flux ratio,
the vertical magnetic field component in the disk is easily determined by the relation
\beq
B_z = \alpha \, 2 \pi G^{1/2} \Sigma,
\label{isopedic}
\eeq
where $\mu \equiv \alpha^{-1}$ is the mass-to-flux ratio in units of the critical 
value for collapse $(2 \pi G^{1/2})^{-1}$ for a thin sheet \citep{Nakano}.
We note that during the isothermal phase of evolution, both the gas pressure and
magnetic pressure gradient terms can be simplified so that their sum 
equals $(1 + 2 \alpha^2)\csq \nabla_p \Sigma$ \citep{Basu,SL,Nakamura}.

If we assume that the material above the thin disk (already assumed to carry
negligible inertia) is also current-free
($\nabla \times \bl{B} = 0$),
then the magnetic field above the disk 
satisfies $\bl{B}(r,\phi,z,t) = \nabla{\Psi}(r,\phi,z,t)$, 
where $\Psi$ is a scalar magnetic potential.
The condition $\nabla \cdot \bl{B}=0$ means that we can determine 
$\Psi$ by solving 
\beq
\nabla^2 \Psi = 0,
\label{laplacemag}
\eeq
subject to the boundary conditions (in the limit $Z \rightarrow 0$)
\beq
\frac{\partial \Psi}{\partial z} =  B_z \: (z = 0), \: \: 
\Psi \rightarrow 0 \: (z \rightarrow \infty). 
\label{magbc}
\eeq
Once a solution is obtained, we calculate
\beq
\bl{B}_p(r,\phi,t) = \nabla_p \Psi(r,\phi,z=0,t).
\label{bp}
\eeq
Equations (\ref{laplacegrav})-(\ref{gp}) and (\ref{laplacemag})-(\ref{bp}) show that there 
is a formal similarity in how 
$\bl{B}_p$ is determined from $B_z$ and $\bl{g}_p$ is determined from $g_z 
(=-2 \pi G \Sigma)$.
The symmetry is even stronger due to our assumption of uniform mass-to-flux ratio
(eq. [\ref{isopedic}]), so that 
\beq
\bl{B}_p = - \frac{\alpha}{G^{1/2}} \bl{g}_p  \Rightarrow \frac{B_z \bl{B}_p}{2 \pi} 
= - \alpha^2 \bl{g}_p.
\label{magtens}
\eeq

Altogether, equations (\ref{cont})-(\ref{mom}), (\ref{halfthick})-(\ref{eos}),
(\ref{gp})-(\ref{isopedic}), and (\ref{magtens}) form a closed system 
of equations for the evolution of the thin disk.

\subsection{Numerical Methods}
The thin-disk equations are solved numerically in polar
coordinates ($r,\phi$) using the method of finite differences with a time-explicit,
operator split solution procedure 
similar to that described by \cite{SN} for their ZEUS-2D code.
Advection is performed using the second-order \cite{vanLeer} scheme. 
The timestep is determined 
according to the usual Courant-Friedrichs-Lewy criterion.
The numerical grid has $128\times 128$ points, which are uniformly spaced in 
the azimuthal direction and logarithmically spaced 
in the $r$-direction, i.e., the radial grid points are chosen to lie between the inner
boundary $r_{\rm in}$ and outer boundary $r_{\rm out}$ so that 
they are uniformly spaced in $u = \log r$. With 128 radial points, the innermost
cell size is 0.56~AU for our typical adopted values $r_{\rm in} = 10$ AU and
$r_{\rm out}= 10^4$ AU.
We introduce a ``sink cell'' at $r <  r_{\rm in} = 10$~AU, which represents 
the central protostar plus some circumstellar disk material, 
and impose a free inflow
inner boundary condition.  We trace the value of the gas surface density in the
sink cell and we assume the formation of a central protostar when
it exceeds the critical density $\Sigma_{\rm cr}$. 
The gas that passes through the sink cell is
then distributed in a $90\%:10\%$ proportion between the protostar (the central point object) 
and the inner circumstellar disk. The main results are insensitive to the details
of mass distribution in the sink cell.  
We assume that the matter is cycled through the inner circumstellar disk and onto
the protostar rapidly enough so that the mass infall
through the sink cell is at least proportional to the mass accretion 
rate onto the protostar.

In our models, the gravitational potential $\Phi$ consists of three parts: 
a central protostar, an inner circumstellar disk, and the cloud core. 
The gravitational potential of the protostar is that of a point mass.
The gravitational potential of the inner circumstellar disk ($0~{\rm AU}<r<10$~AU) 
is computed by decomposing it into a series of twenty concentric circular rings of 
constant density and summing the input from each ring. 
The gravitational potential of an $i$-th ring in the plane of the ring 
is computed using a power series expansion in $r$: 
\begin{equation}
\Phi_{i}(r)=-{GM_{i} \over r} \left( 1+ {a_{i}^2 \over 4 r^2} + {a_{i}^4 
\over 16 r^4} + \ldots \right) \, \, (r>a_{i}),
\end{equation}
where $a_{i}$ is the radius of an $i$-th ring and $M_{i}$ is its mass.
The potential due to the extended mass distribution of the cloud
core (i.e., all the material outside the sink cell) is calculated
as described in \citet[p. 96]{BT}.
A change of variables and use of a numerical grid that is
logarithmically spaced in the $r$-direction allows us to reduce
the integral in equation (\ref{potential}) to a sum that can 
be evaluated by a Fast Fourier Transform technique.

The results of some relevant numerical test problems are presented in
the Appendix.

\subsection{Initial and boundary conditions}
Our model cloud cores are composed
of molecular hydrogen with a $10\%$ admixture of atomic helium 
(so that $m=2.33\, m_{\rm H}$, where $m_{\rm H}$ is the atomic hydrogen mass) 
and are initially
isothermal with $T=10$~K ($c_{\rm s}=0.188$~km~s$^{-1}$).
The initial surface density and angular velocity distributions of cloud
cores are based on the analytic best-fits to numerical models of 
collapsing magnetically supercritical cores \citep{Basu}:
\begin{equation}
\Sigma={r_0 \Sigma_0 \over \sqrt{r^2+r_0^2}},
\label{density}
\end{equation}
\begin{equation}
\Omega=2\Omega_0 \left( {r_0\over r}\right)^2 \left[\sqrt{1+\left({r\over r_0}\right)^2
} -1\right].
\label{omega}
\end{equation}
Here, $\Sigma_0$ and $\Omega_0$ 
are the central surface density and angular velocity,
respectively. 
The above profiles have the property that the specific angular
momentum $K=\Omega\,r^2$ is a linear function of the enclosed mass $M$.
We choose a value  $r_0=\sqrt{2} c_{\rm s}^2/(\pi G \Sigma_0)$, 
so that $r_0$ is comparable to the Jeans length of an isothermal sheet. 
More specifically, we note that the gravitational field of a thin disk with the 
surface density (\ref{density}) is
\beq
g_r = \frac{- 2 \pi G \Sigma_0 r}{(r^2 + r_0^2)^{1/2}
[1 + (1 + r^2/r_0^2)^{1/2}]}\:.
\eeq
For small radii, the ratio of thermal pressure acceleration to
gravity is $a_{\rm T}/|g_r| = \cs^2/\pi G \Sigma_0  r_0$, and is equal to
0.707 for the choice of $r_0$ above. The ratio $a_{\rm T}/|g_r|$
is half this value in the large radius limit.

We introduce a weak nonaxisymmetric perturbation into the initially axisymmetric
density distribution by substituting $r^2$ in equation~(\ref{density}) with
$r^2 ({\cos^2\phi/a^2}+a^2 \sin^2\phi )$. The parameter $a=0.98$ denotes the
cloud oblateness.
Rotation and magnetic fields are introduced in some models through the
parameters $\Omega_0$ and $\alpha$. They are chosen so that 
the cores are initially gravitationally unstable, i.e.,
the ratio of the sum of rotational, magnetic, and thermal energies to the 
magnitude of gravitational energy is less than unity.
We emphasize that
our qualitative results are insensitive to the particular choice of 
initial surface density and angular velocity distributions.

We impose an outer boundary condition such that the gravitationally bound
cloud core has a constant mass and volume. The assumption of a constant mass
is observationally justified by the detection of sharp boundary edges in
the radial gas density distribution of pre-stellar cloud cores \citep{WT,Bacmann}.
Physically, this assumption is justified if the cloud core can dynamically
decouple from the parent diffuse cloud due to a much shorter dynamical timescale 
in the contracting central condensation than in the external region.
A specific example of such a decoupling is the ambipolar-diffusion induced
core formation in magnetically supported clouds \citep[e.g.,][]{Basu95b}.
The assumption of a constant volume implies a constant radius of gravitational
influence of a cloud core within a parent diffuse cloud.

\section{Results}
\label{results}

\subsection{Mass accretion rate in nonrotating cloud cores}
\label{nonrotate}

The behavior of the mass accretion rate onto the protostar plus circumstellar
disk system in spherical {\it finite-mass} models was analytically and numerically
studied by \citet{Henriksen},  
\citet{WW01}, and \citet{VB1}. 
They showed the importance of a finite
mass reservoir as the  cause of an ultimately declining 
mass accretion rate, an effect that is necessarily not part of 
a self-similar solution like that of \cite{Shu77}.
In particular, \citet{VB1} argued that 
a finite mass reservoir and an associated phase of declining mass 
accretion rate and declining bolometric luminosity is necessary to understand the 
Class~I evolutionary phase of protostars.

In this section, we present a prototype cloud core (hereafter, model~1) 
with outer radius $r_{\rm out}=10^4$~AU and mass $M_{\rm c}=1.0~M_\odot$.
It is nonrotating ($\Omega_0=0$) and nonmagnetic ($\alpha=0$). 
The parameters
of this and other models are listed in Table~1. All model cloud cores have
$\Sigma_0=0.12$~g~cm$^{-2}$ (corresponding to a central number density
$n_0=10^6$~cm$^{-3}$), but differ in the
values of other parameters ($\Omega_0,~\alpha,~\gamma$, etc.). 

\begin{figure}
\plotone{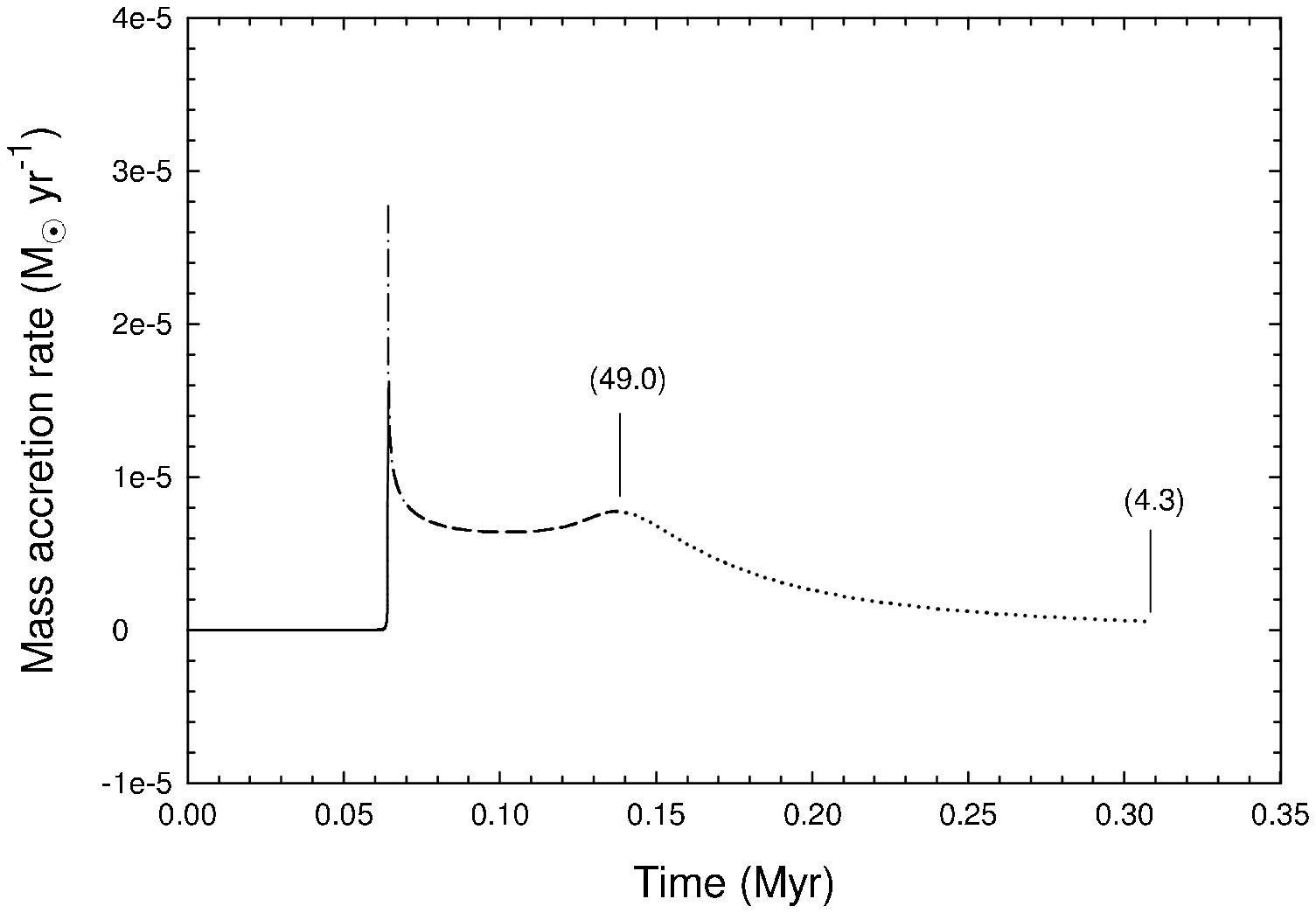}
      \caption{Temporal evolution of the mass accretion rate $\dot{M}$ in the nonrotating
      and nonmagnetized model~1. Three distinct phases in $\dot{M}$ are
      shown by the dash-dotted, dashed, and dotted lines (see explanation
      in the text). The numbers in parentheses indicate the percentage of the total
      cloud core mass left in the envelope at the corresponding evolutionary
      time.   }
         \label{fig1}
\end{figure}

We compute the mass accretion rate $\dot{M} = - 2 \pi r \Sigma v_r$ 
after protostar formation 
at $r = r_{\rm in} = 10$ AU in model~1,
and compare the results with that obtained 
in spherical collapse models by \citet{VB1}.
Figure~\ref{fig1} shows the temporal evolution of $\dot{M}$, which is 
characterized by {\em three distinct phases} after protostar
formation at $t=0.064$~Myr. In the early phase, shown in Figure~\ref{fig1}
by the dash-dotted line, $\dot{M}$ is declining due to a gradient of infall
velocity in the inner region, which is an effect not predicted by 
isothermal similarity solutions. In the intermediate phase that 
follows the initial
decline in $\dot{M}$ and is shown in Figure~\ref{fig1} by the
dashed line, $\dot{M}$ attains a near constant value ($\propto \cs^3/G$)
which is consistent
with the standard theory of \citet{Shu77}. The later phase of 
mass accretion, shown in Figure~\ref{fig1} by the dotted
line, is a result of the finite mass reservoir in our model. 
A shortage of matter developing at the outer edge of a core generates
an inward-propagating rarefaction wave that steepens the radial
gas density profile. When accretion occurs from the mass shells
in the region affected by the rarefaction wave at the time of protostar
formation, a rapid and terminal decline in $\dot{M}$ occurs.
These three phases, calculated in disklike geometry, confirm the general
picture described by \citet{VB1} for spherical 
hydrodynamic collapse. 
However, unlike the spherically symmetric model, there is a small 
rise in $\dot{M}$ between the intermediate and late
phases of mass accretion. This effect is due to the flattened
geometry of the cloud core. The gravitational field at the edge of a
truncated flattened core is greater in magnitude than the corresponding
field in an 
infinite cloud with a low density tail. This is due to the fact that
the net gravitational field acting on any mass shell is a sum of contributions
from both the inner and outer mass shells, unlike in spherical geometry.
The missing effect of an outward pull from mass that would be located outside
the cloud boundary makes the net (inward) gravity increase in magnitude
near the cloud boundary rather than continue the
expected monotonic decline.
The extra inward pull on these mass shells leads to the small
rise in $\dot{M}$ during the latter part of the intermediate phase. 
This effect is accentuated in our model by the use of the
infinitesimally-thin-disk approximation for calculating the gravitational field.
 

Hereafter, we refer to the three phases of accretion as the early-decline
(or early) accretion phase, the intermediate-near-constant (or intermediate)
accretion phase, and the
terminal-decline (or terminal) accretion phase.
We find that sufficiently extended cloud cores 
show all three phases while more compact cores usually do not
experience the intermediate phase. For example, 
the cloud core in model 1 has $r_{\rm out}/r_0\approx 7$, and 
is sufficiently large to 
exhibit a well developed intermediate accretion phase.
This result is qualitatively 
consistent with our earlier results in spherical geometry
\citep {VB1}.

\subsection{Mass accretion rate in rotating cloud cores}
\label{rotate}
In Paper I, we 
demonstrated through high-resolution simulations in the thin-disk
approximation that the collapse of a rotating cloud core and 
self-consistent formation of a protostellar disk leads to
a burst mode of protostellar accretion.
The earlier smooth mode of accretion is reminiscent of the 
collapse of a spherical cloud \citep{VB1} or a nonrotating
disklike cloud as presented in the previous section. 
The later burst mode starts when the protostellar disk forms around 
the protostar. In this mode, the mass accretion is distinguished 
by very short but vigorous accretion bursts, which are interspersed 
within longer periods of quiescent low-rate accretion.

In this section, we investigate the role of rotation 
in determining the amplitude and frequency of mass accretion bursts and implied observable quantities.
We also demonstrate how the two {\it modes} of accretion in rotating cloud
cores are linked with the three {\it phases} of mass accretion in nonrotating
cloud cores. 
The rotational and gravitational energies of the initial 
(axisymmetric) cloud core are calculated here as
\begin{equation}
E_{\rm rot}= 2 \pi \int \limits_{r_{\rm in}}^{r_{\rm
out}} r a_{\rm c} \Sigma \, r \, dr,
\label{rotEn}
\end{equation}
\begin{equation}
E_{\rm grav}= - 2\pi \int \limits_{r_{\rm in}}^{\rm r_{\rm out}} r
g_r \Sigma \, r \, dr,
\label{gravEn}
\end{equation}
respectively, 
where $a_{\rm c} = \Omega^2 r$ is the  
centrifugal acceleration. 

We study in detail a cloud core that has the same parameters as model~1
but also rotates according
to equation~(\ref{omega}). This cloud core (hereafter, model~2) has a 
central angular velocity $\Omega_0=1.5$~km~s$^{-1}$~pc$^{-1}
= 4.86 \times 10^{-14}$ rad~s$^{-1}$, leading
to a ratio of centrifugal acceleration to gravity $a_{\rm c}/|g_r| = 
\Omega_0^2 r_0/\pi G \Sigma_0 = 1.87 \times 10^{-3}$ at small radii
and a ratio of cloud energies   
$\beta= E_{\rm rot}/|E_{\rm grav}|=0.275\%$.
The centrifugal acceleration will never catch up with gravity in 
the inner regions of the cloud core during the prestellar collapse phase 
\citep{Norman,Narita,Basu94}, since both terms scale as $1/r_{\rm m}^3$,
where $r_{\rm m}$ is the radius of a Lagrangian mass shell \citep[see 
discussion in][\S\ 3.1.2]{Basu95a}. However, a centrifugal disk can form after protostar
formation, when the gravity is dominated by the central point mass and
scales as $1/r_{\rm m}^2$.

\begin{figure}
\plotone{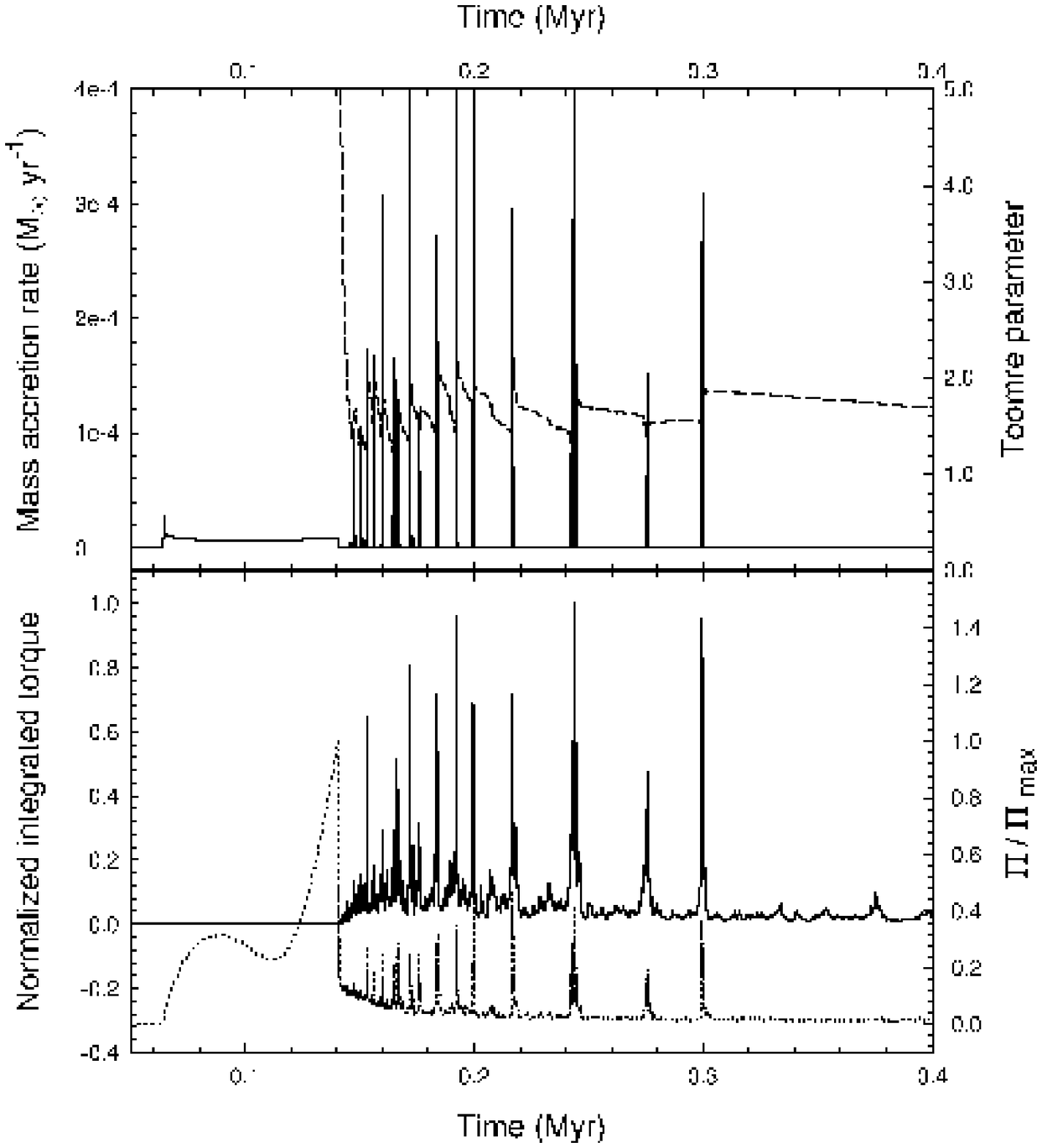}
      \caption{Evolution of model~2. {\em Top}: Temporal evolution of the mass accretion rate 
      ({\em solid line}) and the $Q$-parameter ({\em dashed line}). {\em Bottom}: the normalized 
      integrated torque $\cal{T}/{\cal{T}_{\rm max}}$ ({\em solid line}) and the normalized 
        integrated radial acceleration $\Pi/\Pi_{\rm max}$ ({\em dotted line}).}
         \label{fig2}
\end{figure}

Figure~\ref{fig2} (top) shows the temporal evolution of the mass accretion rate
(solid line) and the Toomre $Q$-parameter (dashed line), whereas 
Figure~\ref{fig2} (bottom)
shows the normalized integrated gravitational torque $\cal{T}$ (solid line) and 
the normalized integrated radial acceleration $\Pi$
(dotted line). 
For our purpose, we are interested in a global property of the disk
(instead of concentrating on local variations), so we
calculate an approximate global value of $Q=\tilde{c}_{\rm s} \Omega 
/\pi G \Sigma$ by averaging $\tilde{c}_{\rm s}$, $\Omega$, and $\Sigma$ 
over all computational cells.
The quantity $\tilde{c}_{\rm s} \equiv \sqrt{d{\cal P}/d\Sigma}$ 
is the effective sound speed. 
We calculate $\cal{T}$ as the sum of the absolute values of the
individual torques $\tau=- m_{\rm c}(r,\phi) \, \partial \Phi/ \partial \phi $
over all computational
cells, and $\Pi$ as the sum of the absolute values of individual
local accelerations $|\partial v_{r}/ \partial r|$ over all cells.
Here, $m_{\rm c}(r,\phi)$ is the gas mass in a cell with the polar coordinates
$(r, \phi)$.
Figure~\ref{fig2} (top) demonstrates that the temporal behavior of $\dot{M}$
after protostar formation at $t \approx 0.06$ Myr
has two distinct modes, as shown in Paper I.
In the earlier smooth mode, the behavior of $\dot{M}$ is very similar to that
of the nonrotating model~1. 
The mass accretion rate stays near $10^{-5}~M_\odot$~yr$^{-1}$,
except for a short period after the protostar formation at $t\approx 0.06$~Myr when $\dot{M}$
is nearly a factor of 3 greater. The later burst mode starts at $t\approx 
0.14$~Myr, when a centrifugal disk forms and the accretion rate abruptly
drops down to a very small value $\dot{M} \la 10^{-7}~M_\odot$~yr$^{-1}$. The 
subsequent evolution is characterized
by very short ($< 100$~yr) but vigorous [$\dot{M}=(1-10)\times 10^{-4}~M_{\odot}$~yr$^{-1}$] accretion bursts, 
which are intervened by longer periods ($\ga 10^3$~yr) of quiescent accretion.
The frequency of the bursts decreases noticeably with time and no bursts
are seen after $t\approx 0.3$~Myr.  

As we demonstrated in Paper I, the accretion bursts occur when the dense 
clumps (henceforth, we often identify them
as protostellar/protoplanetary embryos, or simply as embryos) 
are driven onto the protostar by
the gravitational torques of spiral arms which develop in the disk\footnote{An
animation of the disk evolution for the model presented in Paper I can
be downloaded from \url{http://www.astro.uwo.ca/$\sim$basu/mv.htm}.}.
The temporal evolution of the Toomre parameter $Q$, the integrated 
torque $\cal{T}$, 
and the integrated radial acceleration $\Pi$ proves this scenario.
The $Q$-parameter may serve as an approximate stability criterion since
gas disks are gravitationally unstable to local nonaxisymmetric perturbations 
if $Q\le 1.5-1.7$ \citep{Polyachenko,Nelson,Boss}, while 
$\cal{T}$ may roughly express the efficiency
of angular momentum and mass redistribution by spiral inhomogeneities
in the disk \citep{Tomley}. 
The dashed line in Figure~\ref{fig2} (top) shows that
the Toomre parameter drops below a critical
value just before the burst, indicating the onset of 
gravitational instability.
The integrated torque shown in Figure~\ref{fig2} (bottom) grows before 
the burst and reaches a maximum value at the time of the burst,
suggesting an increase in the efficiency of the radial redistribution of
angular momentum (and mass). 
An increase in the radial redistribution of
mass just before the burst is also implied by the correlation between the
position of bursts in Figure~\ref{fig2} (bottom) and the
maxima in the integrated radial acceleration, since 
the latter may roughly express the degree of centrifugal disbalance
in the disk. 

\begin{figure}
\plotone{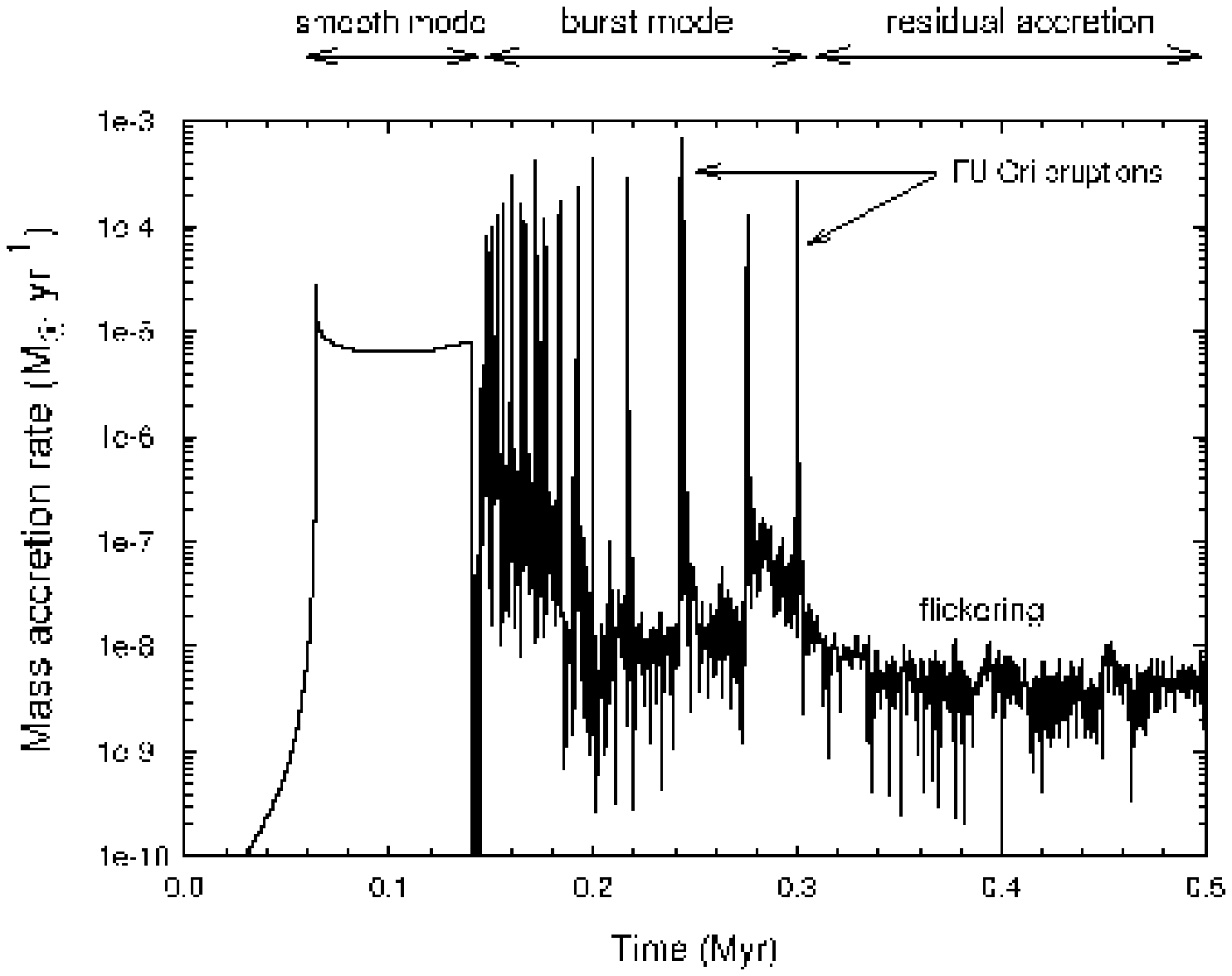}
      \caption{Temporal evolution of the mass accretion rate in
               model~2. A log scale is used
               to emphasize the low-amplitude flickering between the bursts
               and in the late evolutionary stage.}
         \label{fig3}
\end{figure}

\begin{figure}
\centering
\plottwo{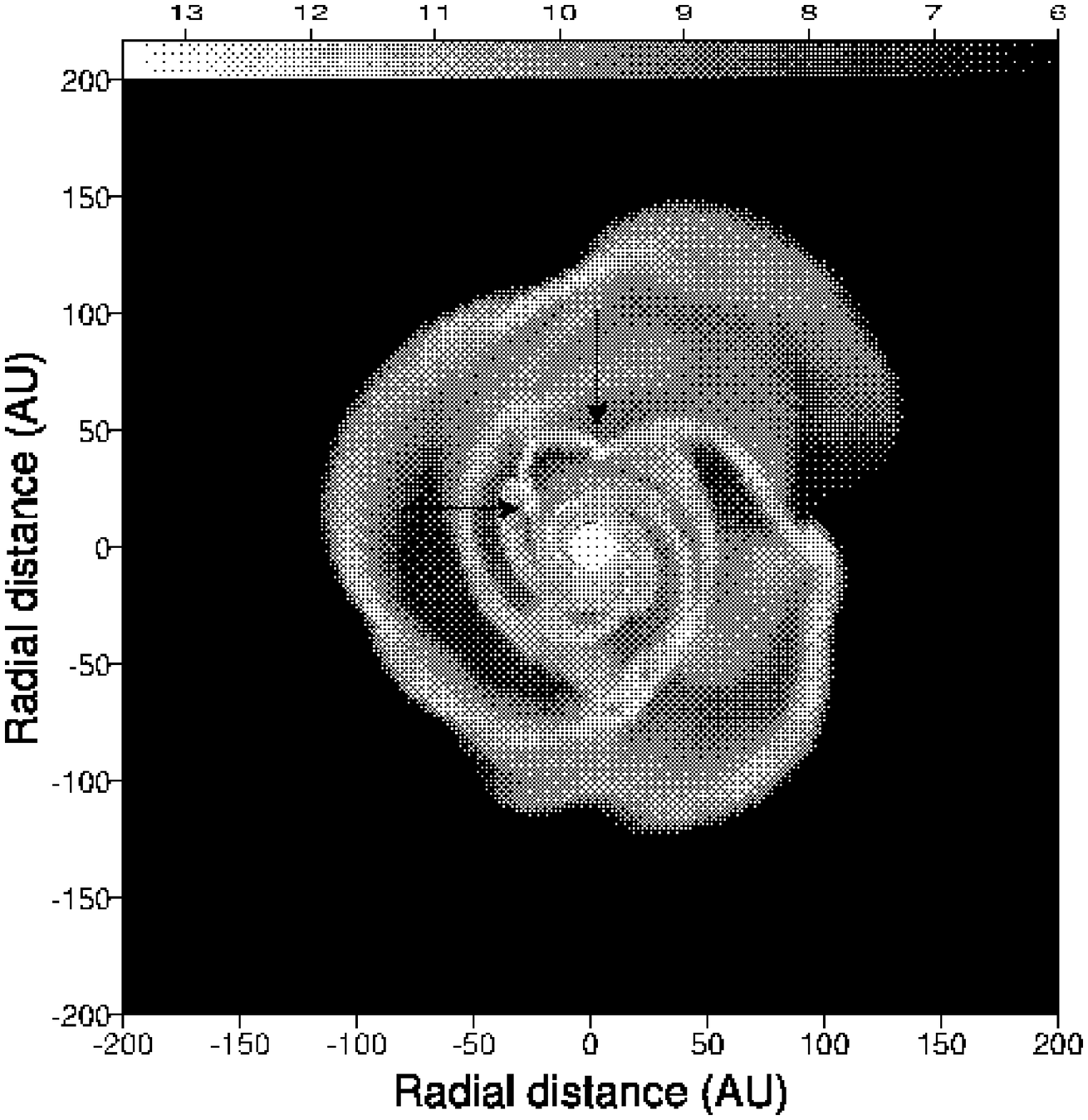}{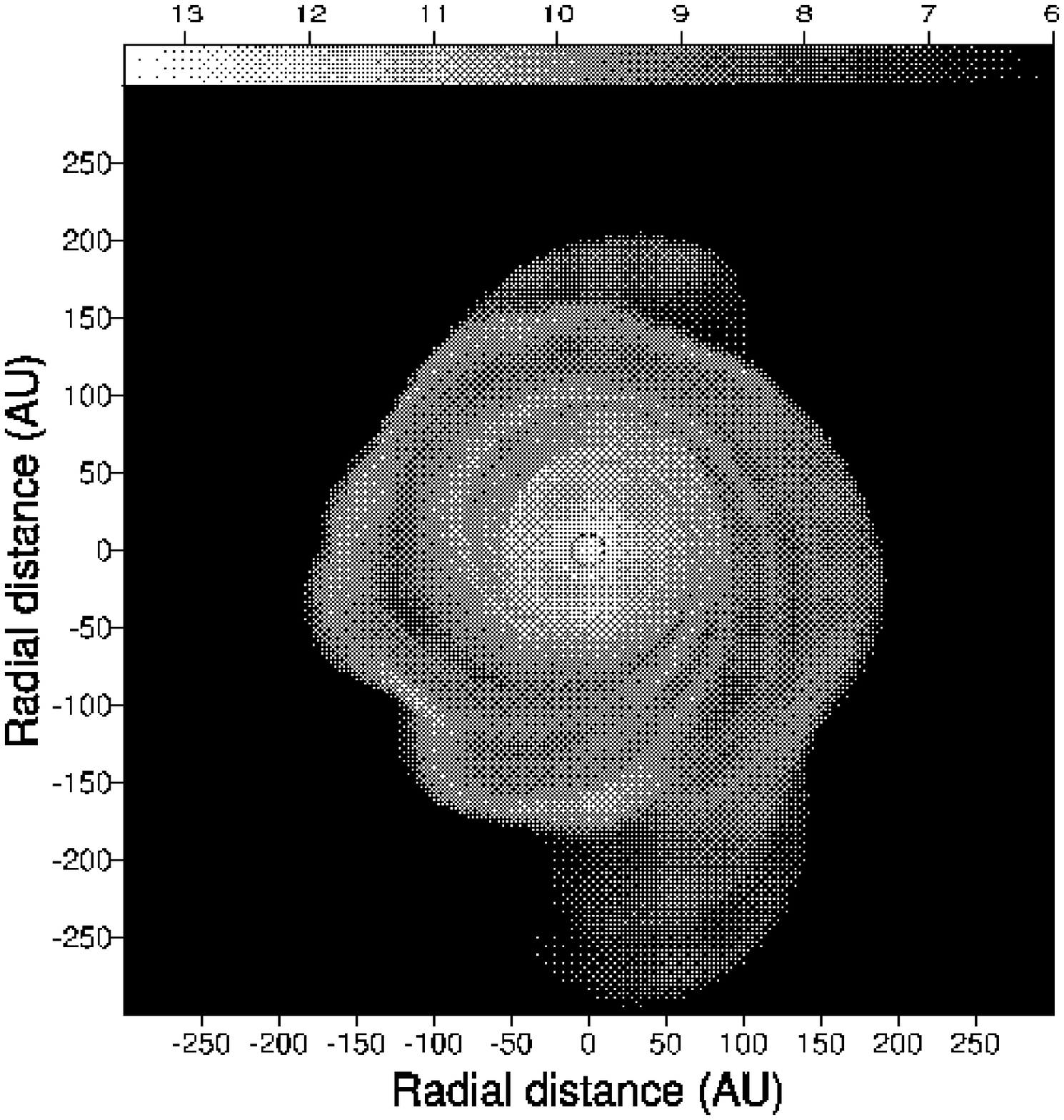}
 \caption{Image of the protostellar disk showing the gas volume
      density distribution in model~2 immediately preceding a mass accretion burst
      ({\em left}) and in the quiescent phase between the bursts ({\em right}).
      The protostellar/protoplanetary embryos with $n \ga
      10^{13}$~cm$^{-3}$ are indicated in the left image with the arrows. The scale bar is
      in cm$^{-3}$. A bright circle in each image represents the protostar
      plus some circumstellar matter.}
 \label{fig4}
\end{figure}

We use a log scale to plot in Figure~\ref{fig3} the mass accretion rate 
obtained in model~2. The log scale emphasizes the low-amplitude flickering in
the burst mode and during the late phase of residual accretion. Note that the flickering
occurs after the protostellar disk forms around the protostar and it
is absent during the smooth mode of accretion. The low-amplitude variations in $\dot{M}$ 
can be caused by gravitational torques of flocculent spiral arms (see Figure~\ref{fig4} 
for an example of spiral structure in the disk).
We note that Figure ~\ref{fig3} bears a remarkable resemblance to the
empirically inferred schematic accretion history of young stars 
presented by \citet[fig. 1.7]{Hartmann}.

To better illustrate the spiral structure and clump formation, 
we run model~2 at a higher resolution of $256\times 256$ grid points 
and plot the distribution of the gas surface density at two different times. 
Figure~\ref{fig4} (left) shows the spiral
structure that immediately precedes a mass accretion burst.
The formation of dense embryos with $n \ga 10^{13}$~cm$^{-3}$ (shown by the arrows) 
within spiral arms is evident. The spiral structure is quite sharp and very
chaotic, which is a consequence of the elevated mass and angular momentum
redistribution and the onset of fragmentation.
In contrast, during the quiescent phase of accretion the protostellar disk typically
has a more uniform spiral pattern, as shown in Figure~\ref{fig4} (right). 
The spiral arms are smoother and more diffuse than in the period immediately
preceding a burst.

\begin{figure}
\epsscale{0.8}
\plotone{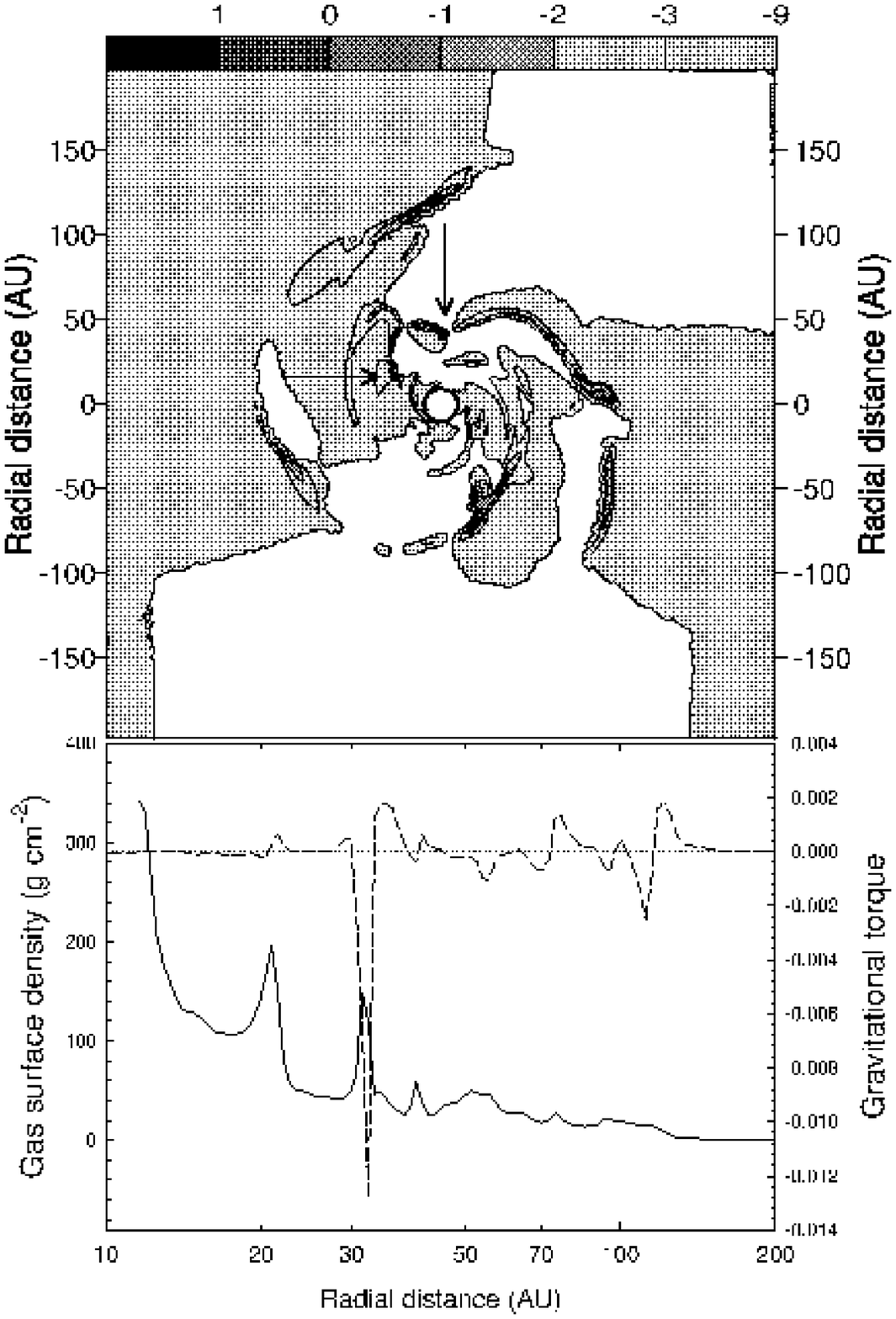}
      \caption{ ({\it Top}) Spatial distribution of the negative gravitational torque
      $\tau$ (by absolute value) corresponding to the gas volume density
      distribution of Fig.~\ref{fig4} ({\it left}).
      The white space is characterized by positive $\tau$
      (not shown in this image). The arrows point to the regions
      of maximum $\tau$ which coincide with the two clumps shown
      in Fig.~\ref{fig4} ({\it left}). The central circle represents the protostar
      plus some circumstellar matter. The scale bar is in dimensionless
      units and the conversion factor is $8.66\times 10^{40}~{\rm dyne}\cdot {\rm
      cm}$. ({\it Bottom}) Azimuthally averaged radial profiles of the surface density ({\it solid line})
      and gravitational torque ({\it dashed line}) obtained from the spatial distribution
      of these quantities shown in Fig.~\ref{fig4} ({\it left}) and the top
      panel of this figure, respectively.}
         \label{fig5}
\end{figure}

While the integrated (by absolute value) gravitational torque $\cal{T}$ 
expresses the efficiency of angular momentum and mass redistribution in the disk
{\it as a whole}, the individual (local) gravitational torques $\tau=-m_{\rm
c}(r,\phi)\, \partial \Phi / 
\partial \phi$ can provide physical insight
into the temporal behavior of {\it local} density inhomogeneities.
In general, the inhomogeneities that are characterized by negative $\tau$ are losing
angular momentum and spiraling onto the protostar, while the inhomogeneities  
that are characterized by positive $\tau$ are gaining angular momentum 
and are moving radially outward.
Figure~\ref{fig5} (top) shows the spatial distribution of $\tau$ computed
at the same evolutionary time as the gas surface density distribution in Figure~\ref{fig4} (left).
Only regions of negative $\tau$ have the absolue values plotted
on a log scale, while the regions of positive $\tau$ are left as white space.
It is evident that the density enhancements are usually (but not always) 
distinguished by the negative gravitational torques. 
The largest negative torques are found at the positions of the two clumps shown
in Figure~\ref{fig5} (top) and Figure~\ref{fig4} (left) by the arrows. 
Note that a large portion in the upper-right part
of Figure~\ref{fig5} (top) is characterized by positive torques, implying a local exchange of angular momentum
between the clumps and this part of the disk. 
In some cases, the radial profiles
of the azimuthally averaged gravitational torque $\tau(r)$  and the azimuthally
averaged gas surface density $\Sigma(r)$ can better capture
the physics behind the burst phenomenon; they are shown in 
Figure~\ref{fig5} (bottom).  It is clearly seen
that the largest negative torque is exerted on the gas density enhancement at $r\approx 32$~AU,
which in fact corresponds to a dense clump shown
in Figure~\ref{fig4} (left) by the horizontal arrow. This clump
is caught in the process of a dramatic loss of its angular momentum and
is currently spiraling down onto the protostar. In return, the gas at
slightly larger radii $r\approx (32-40)$~AU is experiencing the positive gravitational
torque and is currently moving outward. This is a nice example of the 
local angular
momentum and mass redistribution mechanism that ultimately leads to the
burst phenomenon.
We note that the correlation of angular momentum transport with regions of
high surface density fluctuation is qualitatively similar to the 
process described by \citet{Larson84}.

\begin{figure}
\plotone{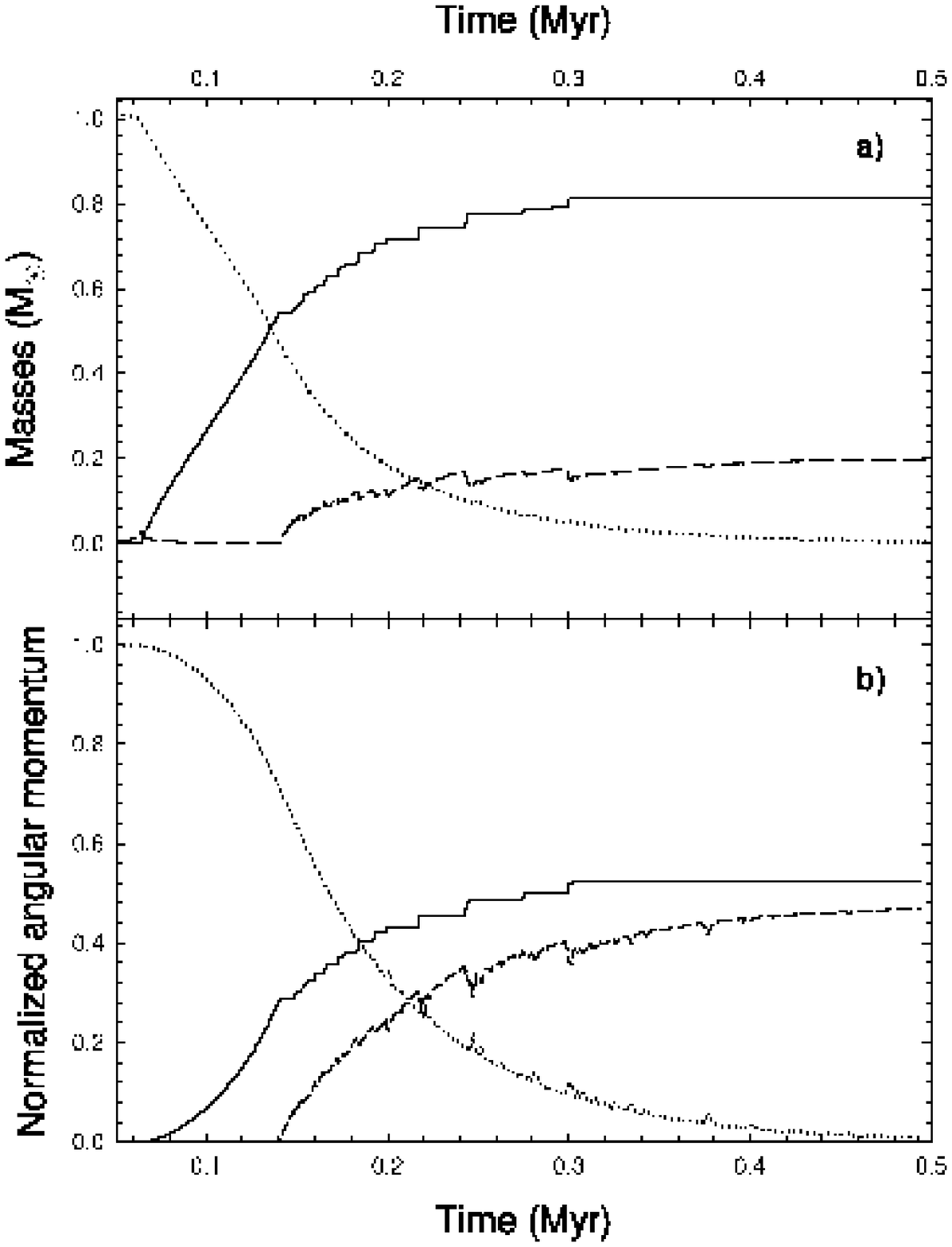}
      \caption{Temporal evolution of ({\bf a}) the protostar mass ({\em solid line}), the 
      envelope mass ({\em dotted line}), and the protostellar disk mass ({\em dashed
      line}) and ({\bf b}) the protostellar angular momentum ({\em solid
      line}), envelope angular momentum ({\em dotted line}), and protostellar
      disk angular momentum ({\em dashed line}) in model~2. 
      Note that the mass of the protostellar disk always stays
      well below that of the protostar.
      By the notion ``protostar'' we refer to the central 10~AU, which comprises the actual
      central object plus some circumstellar disk material.
      }
         \label{fig6}
\end{figure}

Figure~\ref{fig6}a shows the masses contained in the envelope (dotted line),
the protostellar disk (dashed line), and the inner 10~AU (solid line) in model~2.
We note that the inner 10~AU comprises the protostar and some circumstellar
matter. The dynamics of this region is not resolved in our numerical simulations.
However, the mass and total angular momentum contained within that region 
are accurately calculated using the mass and angular momentum fluxes through the inner boundary at
10~AU. Henceforth, we refer to the inner 10~AU as simply
a protostar. The evolution is followed until approximately $99\%$
of the envelope has been absorbed by the protostar and the protostellar disk.
We define the protostellar disk mass as that contained within the inner
200~AU; the material external to this distance is attributed to the envelope.
This gives us upper and lower estimates for the disk and envelope
masses, respectively.
It is evident that every sharp increase in the protostellar mass (associated with 
the infall of embryos) correlates with a corresponding decrease in
the mass of the disk, which {\it remains well below the mass of the protostar
during the evolution}. Approximately $0.01~M_\odot-0.05~M_\odot$ is absorbed during
each mass accretion burst. The accretion bursts cease at $t=0.3$~Myr, when the initial
envelope mass has dropped by roughly $95\%$ and the mass accretion rate
onto the protostellar disk (measured at $r=600$~AU) has decreased
to $6\times 10^{-7}~M_\odot$~yr$^{-1}$. In contrast,
the accretion rate is $7.5\times 10^{-6}~M_\odot$~yr$^{-1}$
at the beginning ($t=0.15$~Myr) of the burst mode of accretion. 
Figure~\ref{fig6}b shows the normalized total angular momenta of the envelope (dotted line), 
the protostellar disk (dashed line), and
the protostar (solid line), respectively, in model~2.
We calculate the total angular momentum of the disk as the sum of the individual angular momenta
$m_{\rm c} \Omega \, r^2$ in the inner 200~AU; the individual angular momenta
external to this distance are attributed to the total angular momentum
of the envelope.
The comparison of the top and bottom panels in Figure~\ref{fig6} reveals
a striking difference in the temporal evolution of masses and total angular momenta 
of the protostar and protostellar disk. At $t=0.5$~Myr, when the envelope has lost more
than $99\%$ of its mass and angular momentum, the protostar has gained almost
80\% of the total mass but only $55\%$ of the total angular momentum. 
In contrast, the protostellar disk has gained at the same time only
$20\%$ of the total mass but almost $45\%$ of the total angular momentum.
This is a powerful confirmation of the angular momentum redistribution in the protostellar disk,
which we attribute to the gravitational torques of spiral arms. 
The fact that the protostar has gained approximately $55\%$ of the total angular momentum after
0.5~Myr may seem excessive. However, we note that we do not resolve the
protostar itself. This amount of angular momentum is in fact contained
in the inner 10~AU, which includes not only the protostar but
some circumstellar matter as well. 
The distribution of angular momentum within the inner 10~AU remains beyond
the scope of our present calculations.
We further note that the thin-disk approximation does not take into account
the powerful mechanism of angular momentum loss by protostellar winds,
which would ultimately take away a considerable portion of the angular momentum in the
inner 10~AU.
Finally, we want to stress that the sum of all masses and angular momenta in Figure~\ref{fig6} remain
practically constant during the evolution, indicating an excellent global conservation of mass
and angular momentum in our numerical simulation.

\begin{figure}
\plotone{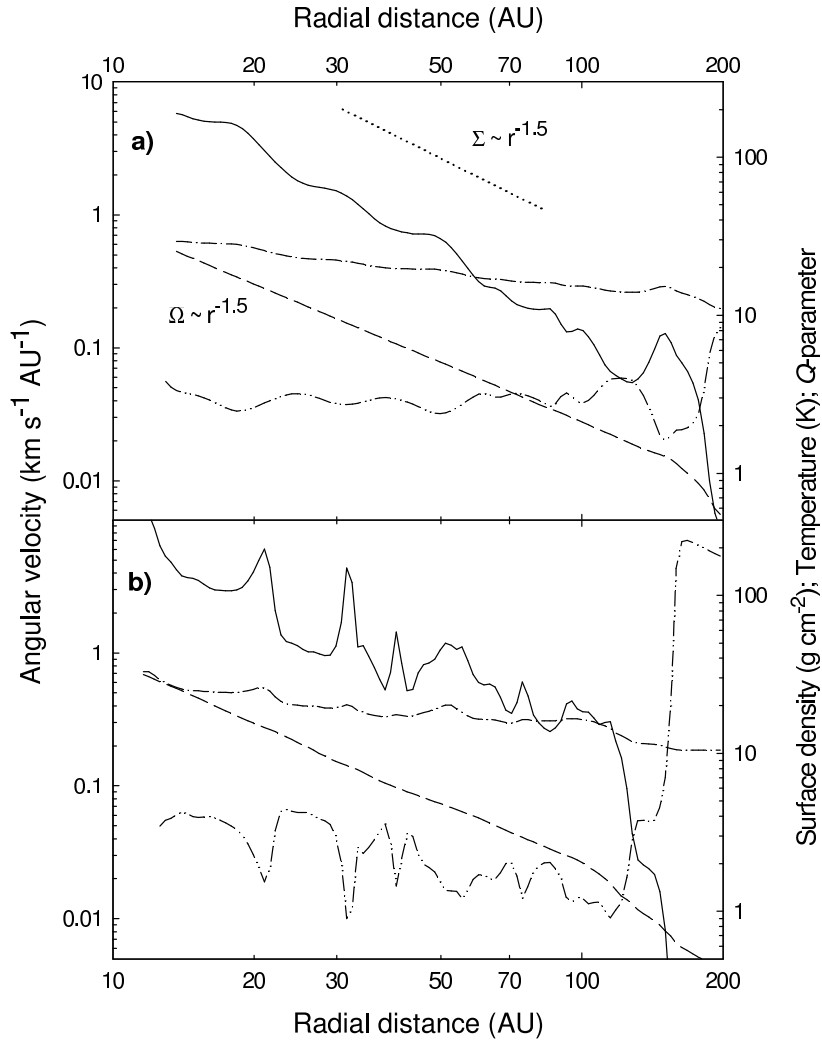}
      \caption{Radial profiles of surface density ({\it solid lines}), angular
      velocity  ({\it dashed lines}), temperature ({\it dash-dotted lines}),
      and $Q$-parameter (dash-double-dotted lines) corresponding
      to ({\bf a}) the gas distribution in the quiescent phase shown in Fig.~\ref{fig4} ({\it right})
      and ({\bf b}) to the gas distribution in the burst phase shown in Fig.~\ref{fig4} ({\it left}).
      Note that the rotation is nearly Keplerian in both cases.
      }
         \label{fig7}
\end{figure}

In order to get a better understanding of 
the physical conditions in the disk, we plot in
Figure~\ref{fig7}a and Figure~\ref{fig7}b the azimuthally averaged radial profiles
of surface density $\Sigma$ (solid lines), angular velocity $\Omega$ (dashed lines), temperature
$T$ (dash-dotted lines), and $Q$-parameter (dash-double-dotted lines) 
in the quiescent and burst phases, respectively. 
More specifically, the radial profiles in Figure~\ref{fig7}a
correspond to the gas distribution shown in Figure~\ref{fig4} (right), while
the radial profiles in Figure~\ref{fig7}b are obtained from the gas distribution
of Figure~\ref{fig4} (left). 
The radial profiles of $\Omega$ and $T$ are similar during the burst and
quiescent phases. The disk rotation is nearly Keplerian, $\Omega \propto r^{-1.5}$, 
as can be expected from the ratio of
disk to protostar masses in Figure~\ref{fig6}. The radial temperature profile
shows a mild increase towards the center from 10~K at 200~AU to approximately
30~K at 15~AU. The biggest differences are seen in the radial surface density distribution
-- the quiescent phase is characterized by a much smoother profile with
$\Sigma \propto r^{-1.5}$. The radial surface density profile in the burst
phase has multiple peaks, some of which correspond to the forming protoplanetary
embryos. The 
profiles in both the burst and quiescent phases have sharply defined boundaries
at approximately 200~AU where the disk merges with the infalling envelope.
Since the radial profiles of $\Omega$ and $T$ are quite similar in the
burst and quiescent phases, the radial distribution of the $Q$-parameter is mostly
determined by the radial variations in $\Sigma$.
As a consequence, the $Q$-profile is much smoother and its values are 
noticeably larger in the quiescent phase than
in the burst phase. During the burst phase, 
both the peaks in $\Sigma$ within the inner disk ($<40$~AU) and most of the outer disk 
are characterized by $Q\approx 1.0-1.5$. During the quiescent phase, $Q$
falls in the range $2.5-3.0$ and never goes below 2.0. 
The existence of gravitational instability
is usually not expected at such high values of $Q$.
Nevertheless, a weak spiral structure can be clearly seen in Figure~\ref{fig4} (right),
which shows the gas volume density distribution during the quiescent phase. This
indicates that once the spiral structure is generated during the burst
phase, there exists an amplification mechanism (or mechanisms) that
sustains at least a low level of this structure for a substantially
long time.
Strong observational evidence for the existence of spiral structure in the
several-million-year-old disks around AB Aurigae \citep{Fukagawa} and HD~100546 \citep{Grady} 
tend to support this conjecture.

\subsubsection{The effect of different rotational energy}

\begin{figure}
\epsscale{.80}
\plotone{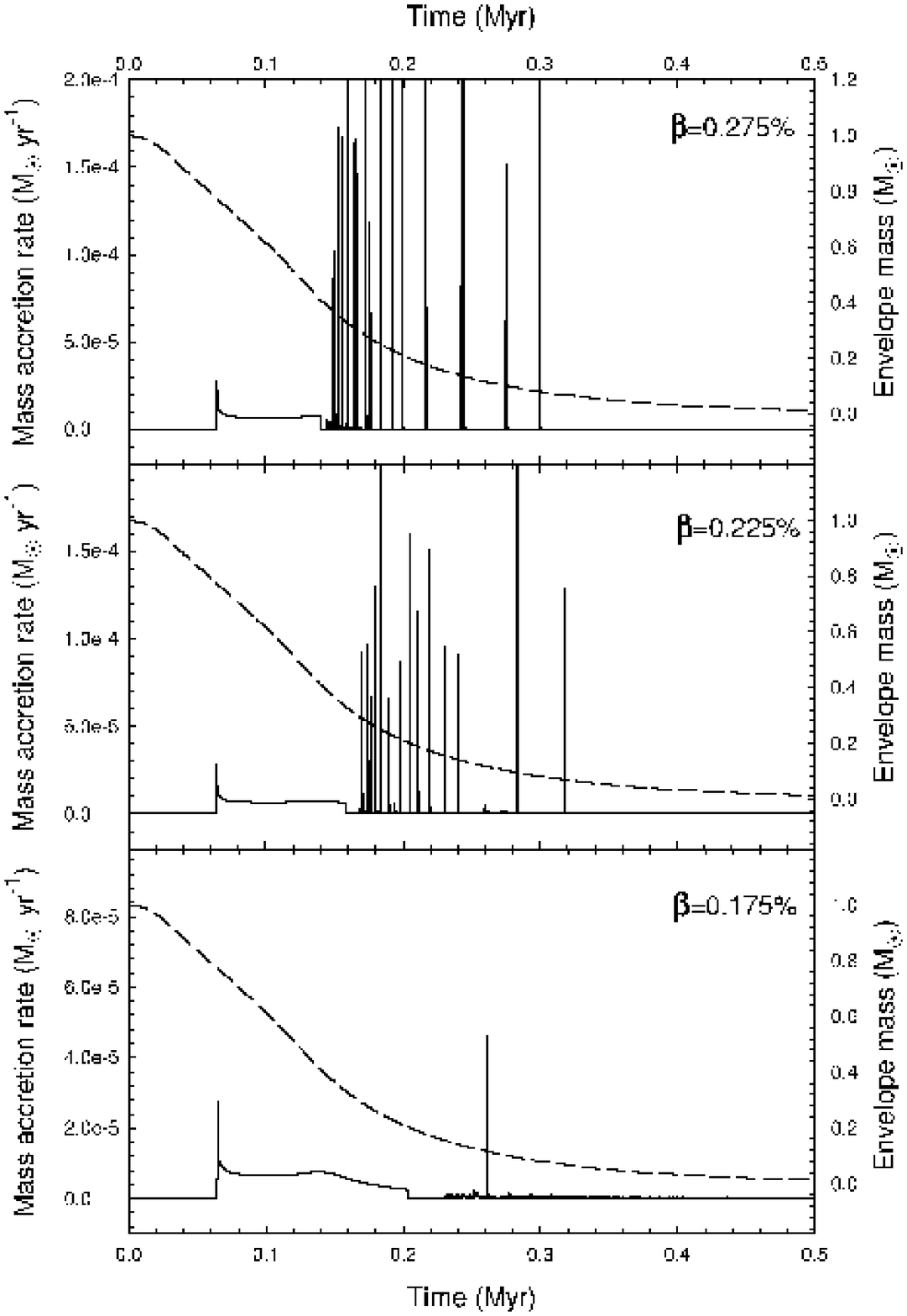}
      \caption{Temporal evolution of the mass accretion rate ({\em solid lines}) and
        envelope mass ({\em dashed lines}) for models with different values of $\beta$. 
        Model~2 ({\em top panel}), model~3 ({\em middle panel}), and model~4 ({\em bottom panel}) 
        have values of $\beta$ as indicated in each panel.}
         \label{fig8}
\end{figure}

As we have shown, rotation introduces a qualitatively
distinct mode into the evolution of mass accretion onto the protostar: a burst
mode. In this subsection, we investigate the effect that the different 
initial rotational
energies of cloud cores may have on the frequency, number, and amplitude of mass accretion bursts.
For comparison with model 2, we present model 3 and model 4, which have different 
initial rotation energies than model 2 but are otherwise identical.
The parameters of these models are listed in Table~\ref{table1}. Since
all three models have the same initial gravitational energy, it is useful
to differentiate the models by $\beta$, the ratio of magnitudes of initial rotational
and gravitational energy.
The solid lines in the top, middle, and bottom panels of Figure~\ref{fig8} show the temporal evolution 
of the mass accretion rate obtained in model~2, model~3, and model~4, respectively.
In each figure, the dashed lines show the instantaneous envelope mass.
A decrease in $\beta$ has a profound effect on the 
bursts. The $\beta=0.225\%$ model~3 has noticeably fewer and less 
vigorous bursts.
A further decrease to $\beta=0.175\%$ (model~4) almost 
eliminates the bursts.

This tendency can be understood if we consider the three phases of mass accretion 
onto the protostar in the case of nonrotating cloud cores. We have
emphasized in \S~\ref{rotate} that the temporal behavior of $\dot{M}$ is very similar in
rotating and nonrotating models before the formation of the protostellar
disk. In the $\beta=0.275\%$ model~2, 
the disk forms in the intermediate accretion phase when the infall rate
onto the disk\footnote{We note that $\dot{M}$ is a slowly varying
function of radius, at least in the inner $(1-2)\times 10^3$~AU.} is near-constant
and equals $\approx 7.5\times
10^{-6}~M_\odot$~yr$^{-1}$. The envelope at the time of the protostellar
disk formation is massive enough ($\approx 45\%$ of the total mass) to
continuously supply the disk with matter for $0.1-0.2$ Myr. 
With an infall rate of $\la 7.5\times
10^{-6}~M_\odot$~yr$^{-1}$ onto the disk, it takes $\ga 10^3$~yr for the 
disk to accumulate $\approx 0.01~M_\odot$. {\it The frequency of the mass accretion 
bursts decreases as the infall rate onto the disk
diminishes during the evolution.} 
In the $\beta=0.225\%$ model~3, the disk forms just after the terminal
accretion phase has begun, and the frequency of bursts is somewhat less
than in model~2. The accretion rate
$\dot{M}$ exceeds $2\times 10^{-4}~M_\odot$~yr$^{-1}$ in only two episodes 
as compared to eight episodes in the $\beta=0.275\%$ model~2. 
Furthermore, in 
the $\beta=0.175\%$ model~4, the protostellar disk forms
late in the terminal accretion phase, i.e., when 
most of the total envelope mass has already
been accreted by the protostar and protostellar disk. Consequently, 
the infall rate onto
the disk is low ($\approx 2\times 10^{-6}~M_\odot$~yr$^{-1}$) and quickly 
decreases further.
The protostellar disk stays near the borderline of stability and provides
a relatively smooth accretion of matter onto the protostar, punctuated 
by only a single mass accretion burst.

The size of a cloud core also affects the frequency of bursts since it
determines the duration of the intermediate (near-constant)
accretion phase. The larger a cloud, the longer the duration of the intermediate
phase since the inward propagating rarefaction wave from the outer boundary 
takes a longer time to affect the protostar.
The centrifugal radius $r_{\rm cf}$ of a gas parcel located initially
at a distance $r$ 
can be estimated by assuming that 
all mass inside $r$ is concentrated in a central
point source of mass $M$. 
In this case $r_{\rm cf}= K^2/G M$ where $K$ is the specific
angular momentum of the gas parcel. Consequently,
two clouds with identical density and rotation profiles 
but different sizes will form the protostellar disk 
at nearly the same time but in different accretion phases. 
The larger cloud will have a longer intermediate phase of accretion onto the
disk, which will drive more vigorous burst activity.

\begin{figure}
\plotone{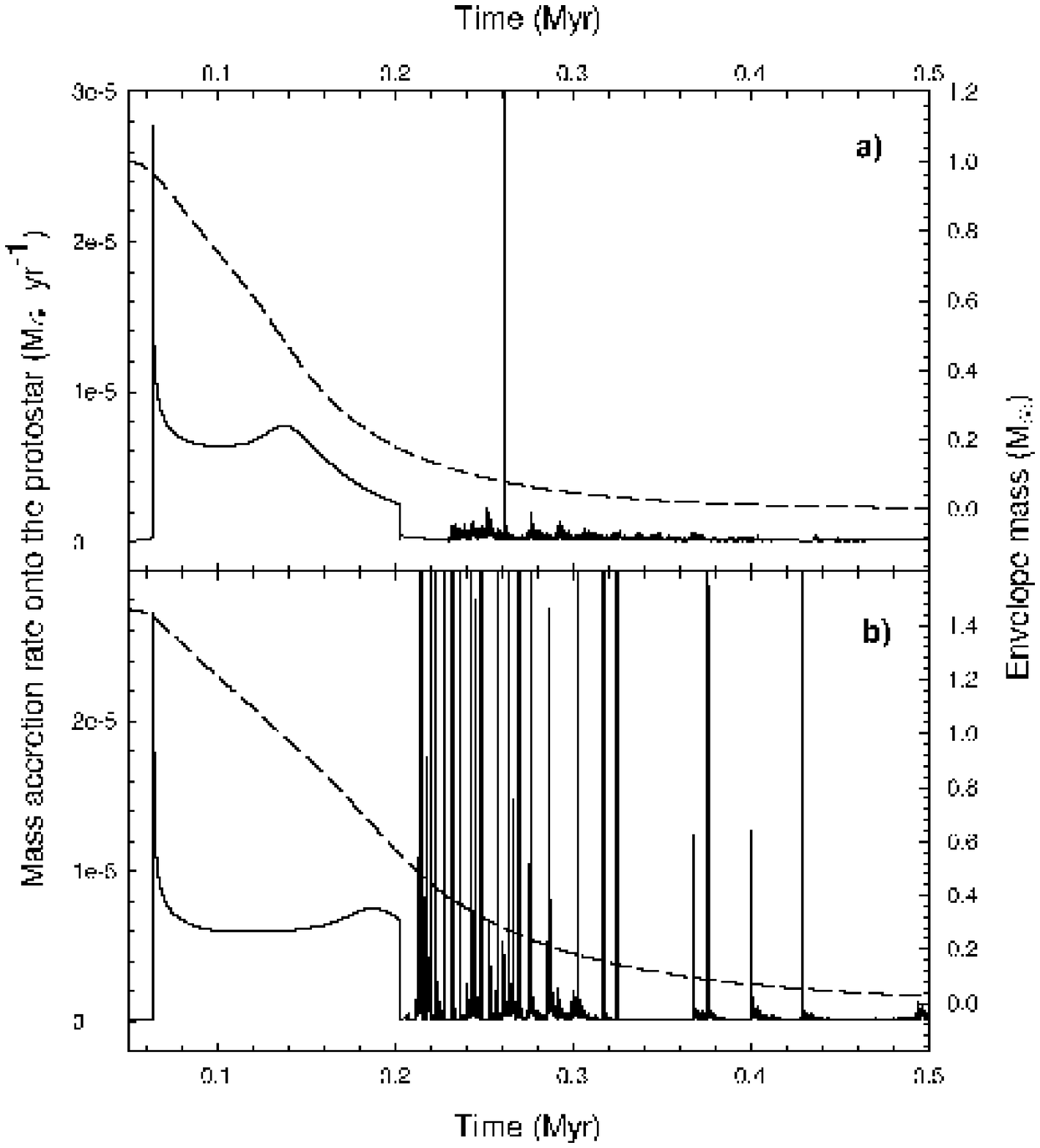}
      \caption{Temporal evolution of the mass accretion rate
      ({\em solid lines}) and envelope mass ({\em dashed lines}) for ($a$) model~4, 
         and ($b$) model~5. The model 5 cloud core has the same initial central rotation rate
        as model~4 but has a greater mass.} 
         \label{fig9}
\end{figure}

Figures~\ref{fig9}a and \ref{fig9}b
show the temporal behavior of the mass accretion rate onto the protostar
(the solid lines) and the corresponding envelope masses (the dashed lines) in model~4 and
model~5, respectively. Both models have identical parameters except that the size of the cloud core
in model~5 is larger than in model~4. More specifically, model~4 has
$r_{\rm out}=10^4$~AU (and $M_{\rm c}=1~M_\odot$), whereas model~5 has $r_{\rm
out}=1.4\times 10^4$~AU (and $M_{\rm c}=1.5~M_\odot$). 
In both models, the protostellar disk forms at approximately 0.2~Myr. 
However, model~5 has just entered the terminal accretion
phase and still has a fairly massive envelope ($0.5~M_\odot$), while model~4 is in the late part of the 
terminal accretion phase and has little mass left in the envelope
($\la 0.1 M_\odot$).
As a result, model~5 shows multiple mass accretion bursts due to a high
mass infall rate ($\la 7\times 10^{-6}~M_\odot$~yr$^{-1}$) 
onto the protostellar disk, whereas model~4 develops a burst only once.

\subsection{Effect of a frozen-in magnetic field in rotating cloud cores}

We include the effect of a magnetic field through the assumptions of
flux-freezing and a spatially uniform mass-to-flux ratio $\Sigma/B_z$,
i.e., the quantity $\alpha$ in equation (\ref{isopedic}) is constant.
In this case, the magnetic tension acts as a simple dilution of gravity
and the magnetic pressure is a multiple of the gas pressure, as discussed in
\S\ \ref{model}.  
This implies that the critical value of the Toomre $Q$-parameter 
in our magnetized disk models should be lower than in 
nonmagnetized disks.

\begin{figure}
\plotone{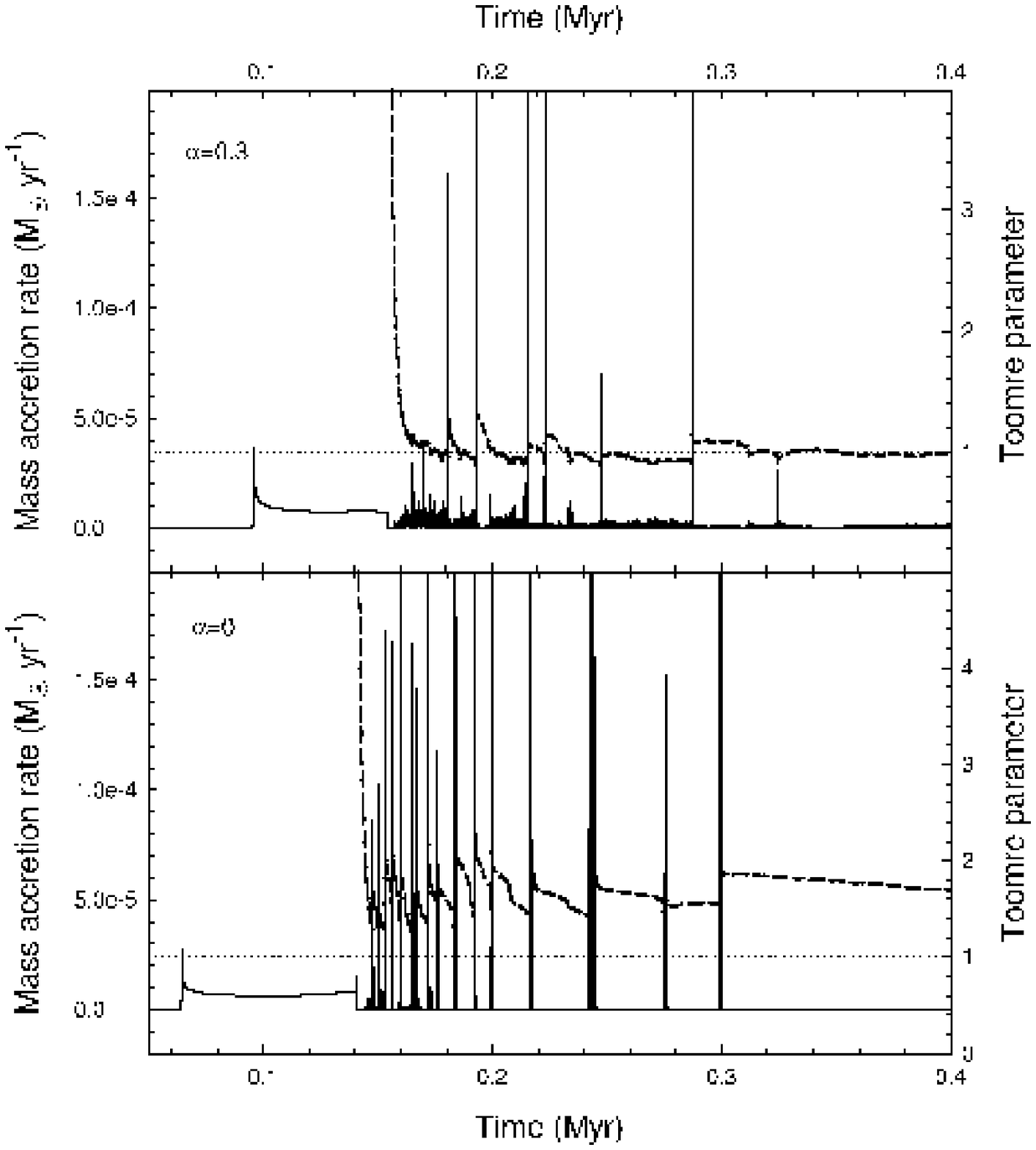}
      \caption{Temporal behavior of the mass accretion rate ({\em solid lines}) and
      the $Q$-parameter ({\em dashed lines}). Shown are model~6 ({\em top}), which has
      $\alpha=0.3$, 
      and model~2 ({\em bottom}), which has $\alpha=0$. The dotted line denotes $Q=1$
        and demonstrates that the magnetized model~6 needs to attain a lower value of
        $Q$ for instability to occur than does the unmagnetized model~2.}
         \label{fig10}
\end{figure}

We simulate a magnetized cloud core 
with $\alpha=0.3$, hereafter called model ~6, which is otherwise identical to the
nonmagnetized model~2. The ratio of magnetic acceleration to gravity
is $a_{\rm M}/|g_r| = \alpha^2(1 + 2\, a_{\rm T}/|g_r|)$, which
equals 0.217 at small radii for $\alpha = 0.3$. Figure~\ref{fig10} shows the
temporal evolution of the mass accretion rate (solid lines) and the
$Q$-parameter (dashed lines) in  
model~6 (the top panel), and in model~2 (the bottom panel). 
The dotted line draws the theoretical borderline
of stability of nonmagnetized gas disks to {\it axisymmetric} perturbations
($Q=1$). 
It is evident that an increase in the magnetization
of the cloud cores delays the formation of the protostar and
moderates the burst activity. 

Can the moderated burst activity in the magnetized model be understood 
in terms of the remaining envelope mass when the disk forms, as in the 
case of models with different levels of rotation?
Comparing the two models, we find that they have comparable envelope
masses at the time of disk formation. In fact, the magnetized model
has a slightly more massive envelope ($0.52~M_\odot$) than does the nonmagnetized
model ($0.45~M_\odot$). This is due to the fact that magnetized disks can
support larger envelopes than their nonmagnetized counterparts. 
Therefore, the decreased burst activity in the magnetized disk cannot be
attributed to a smaller envelope and mass accretion rate.
Instead, it is an increased intrinsic stability  
of magnetized disks that is the key factor that moderates 
the mass accretion bursts.

The temporal behavior of the Toomre $Q$-parameter in Figure~\ref{fig10}
proves our expectations that the magnetized disks are in general more 
stable than their nonmagnetized counterparts. 
In the magnetized model~6,
the bursts occur at $Q \approx 0.9$, whereas in the nonmagnetized model~2,
the bursts occur when $Q \approx 1.4$.
The latter is in approximate agreement with previous results on
the instability of nonmagnetized disks to nonaxisymmetric perturbations
\citep[e.g.,][]{Polyachenko,Nelson,Boss}. 
The magnetized disks need to attain a lower value of $Q$
in order to be destabilized, fragment into dense clumps, and produce mass
accretion bursts.

\subsection{The isothermal-adiabatic 
transition and ratio of specific heats}
\label{freeparam}
The thermal evolution of a protostellar disk is determined by equation (\ref{tempr}).
The transition between the isothermal and adiabatic regimes is controlled by the critical
density $\Sigma_{\rm cr}$, which can be determined by 
comparison with radiation 
transfer simulations of gravitational collapse.  

\begin{figure}
\plotone{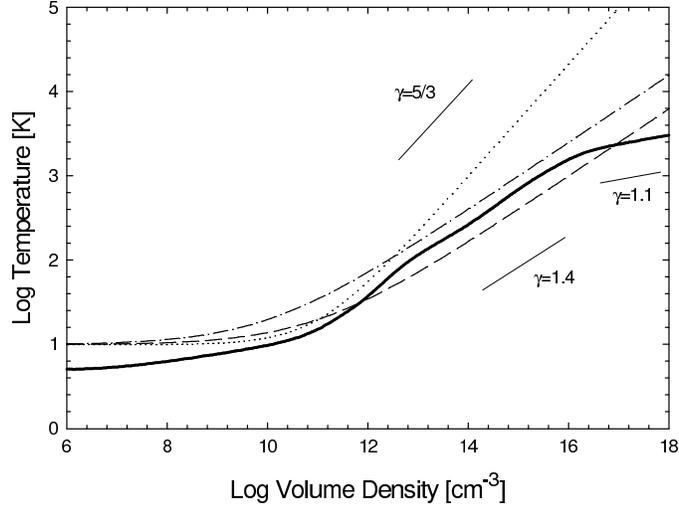}
      \caption{Comparison of the density-temperature relation~(\ref{tempr}) 
      for $\Sigma_{\rm cr}=36.2$~g~cm$^{-2}$ and $\gamma=1.4$ ({\em dashed line}),
      $\Sigma_{\rm cr}=11.6$~g~cm$^{-2}$ and $\gamma=1.4$ ({\em dash-dotted
      line}), and $\Sigma_{\rm cr}=36.2$~g~cm$^{-2}$ and $\gamma=5/3$ ({\em dotted
      line})
      with the calculated density-temperature relation ({\em thick solid line}) of 
      \citet{Masunaga} from radiation transfer simulations 
      of the gravitational collapse of a molecular cloud core.}
         \label{fig11}
\end{figure}

Figure~\ref{fig11} compares our density-temperature relation
with that calculated by 
\citet[hereafter MI]{Masunaga} using spherically symmetric 
frequency-dependent radiation transfer simulations.
The dashed line shows equation (\ref{tempr}) for $\Sigma_{\rm cr}=36.2$~g~cm$^{-2}$
($n_{\rm cr}=10^{11}$~cm$^{-3}$), whereas the dash-dotted line shows it
for $\Sigma_{\rm cr}=11.6$~g~cm$^{-2}$ ($n_{\rm cr}=10^{10}$~cm$^{-3}$).
The thick solid line gives the density-temperature dependence as derived
by MI.  It is evident that $n_{\rm cr}=10^{10}$~cm$^{-3}$
and $n_{\rm cr}=10^{11}$~cm$^{-3}$ may be considered to be
limiting values for the critical density, since most of MI's density-temperature
curve lies between the curves specified
by these critical values. An obvious failure of the polytropic curves to
embrace MI's curve at $n<10^{11}$~cm$^{-3}$ is caused by a lower
initial temperature (5~K) in MI's simulations than our assumption of 10~K.
We note that MI's density-temperature dependence is derived
for spherical collapse and it may differ in case of flattened cloud cores.

\begin{figure}
\plotone{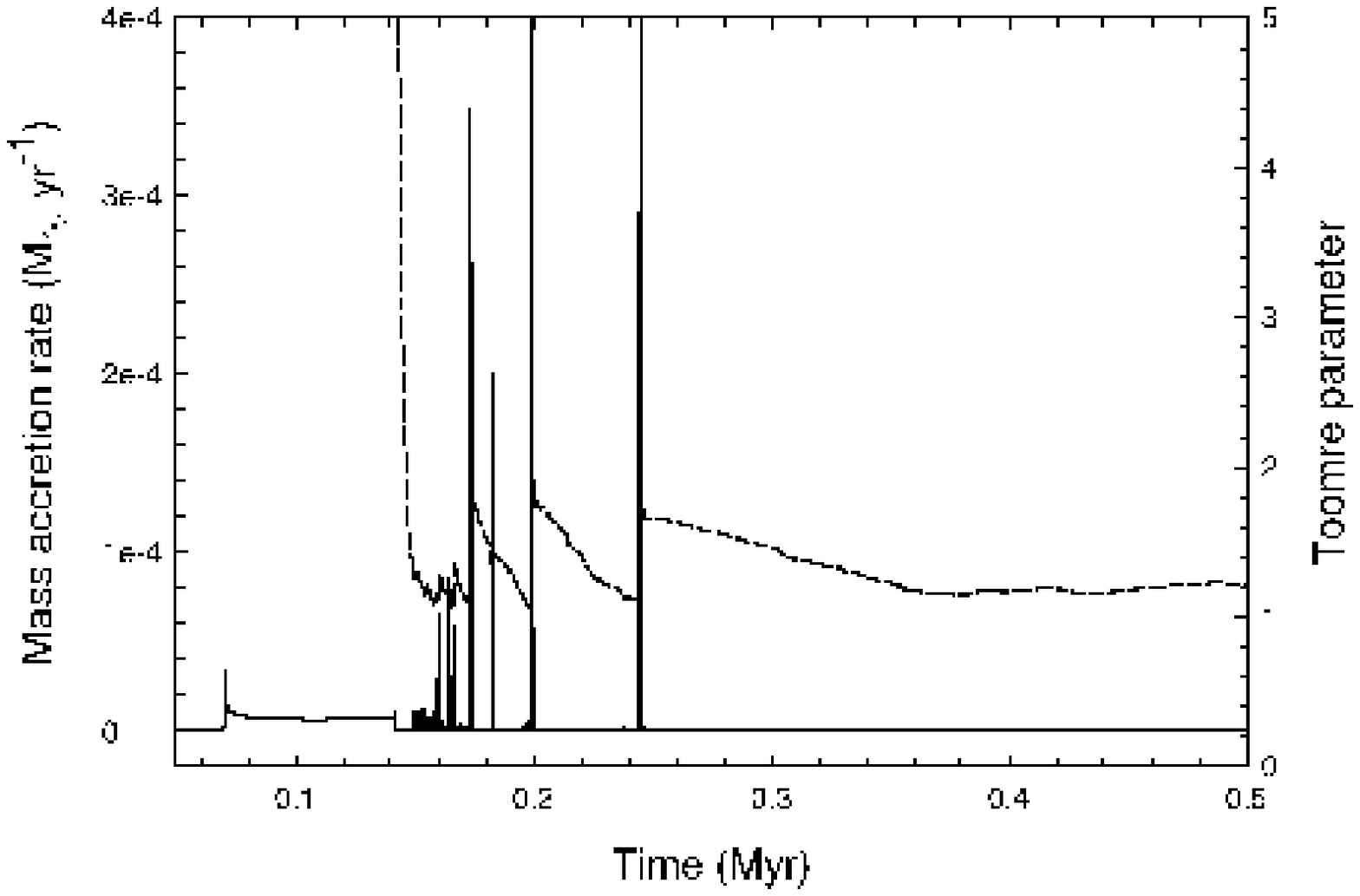}
      \caption{Temporal evolution of the mass accretion rate ({\em solid
      line}) and $Q$-parameter ({\em dashed line}) in model~7, which has the same
      parameters as model~2 but a lower value
      of $\Sigma_{\rm cr}$ ($=11.6$~g~cm$^{-2}$).}
         \label{fig12}
\end{figure}

In the models already presented, we have used $\Sigma_{\rm cr}=36.2$~g~cm$^{-2}$.
For a smaller value of $\Sigma_{\rm cr}$, the disk should be (on average)
hotter and less susceptible to the development of gravitational instability
and formation of protostellar/protoplanetary embryos. Consequently, 
the frequency and number of the mass accretion bursts is expected to decrease.
Indeed, the solid line in Figure~\ref{fig12} shows the evolution of 
$\dot{M}$ in model~7, which
has $\Sigma_{\rm cr}=11.6$~g~cm$^{-2}$ ($n_{\rm cr}=10^{10}$~cm$^{-3}$). 
The parameters of model~7 are otherwise identical to the previously studied model~2 (see Table~1).
A comparison of Figures~\ref{fig2} and \ref{fig12} reveals that 
the number of bursts has decreased by roughly a factor of 3. 
We conclude that a decrease in $\Sigma_{\rm cr}$ and an associated increase
in the overall disk temperature, within realistic bounds, can moderate the 
protostellar/protoplanetary
embryo formation (and related burst activity) but not completely suppress it.

The ratio of specific heats $\gamma$ that enters the polytropic law~(\ref{eos})
will regulate the thermal evolution of the protostellar disk in the optically thick
regime.
The cloud cores consist mostly of molecular hydrogen, which is a diatomic
gas with $\gamma=7/5 = 1.4$. The dashed and dash-dotted lines in Figure~\ref{fig11} demonstrate
that $\gamma=1.4$ reproduces, for the most part, the slope of the 
density-temperature relation from
the simulations of MI. However, if the gas temperature is low, 
neither rotational nor vibrational modes of molecular hydrogen are excited
and $\gamma$ becomes equal to 5/3. 

\begin{figure}
\plotone{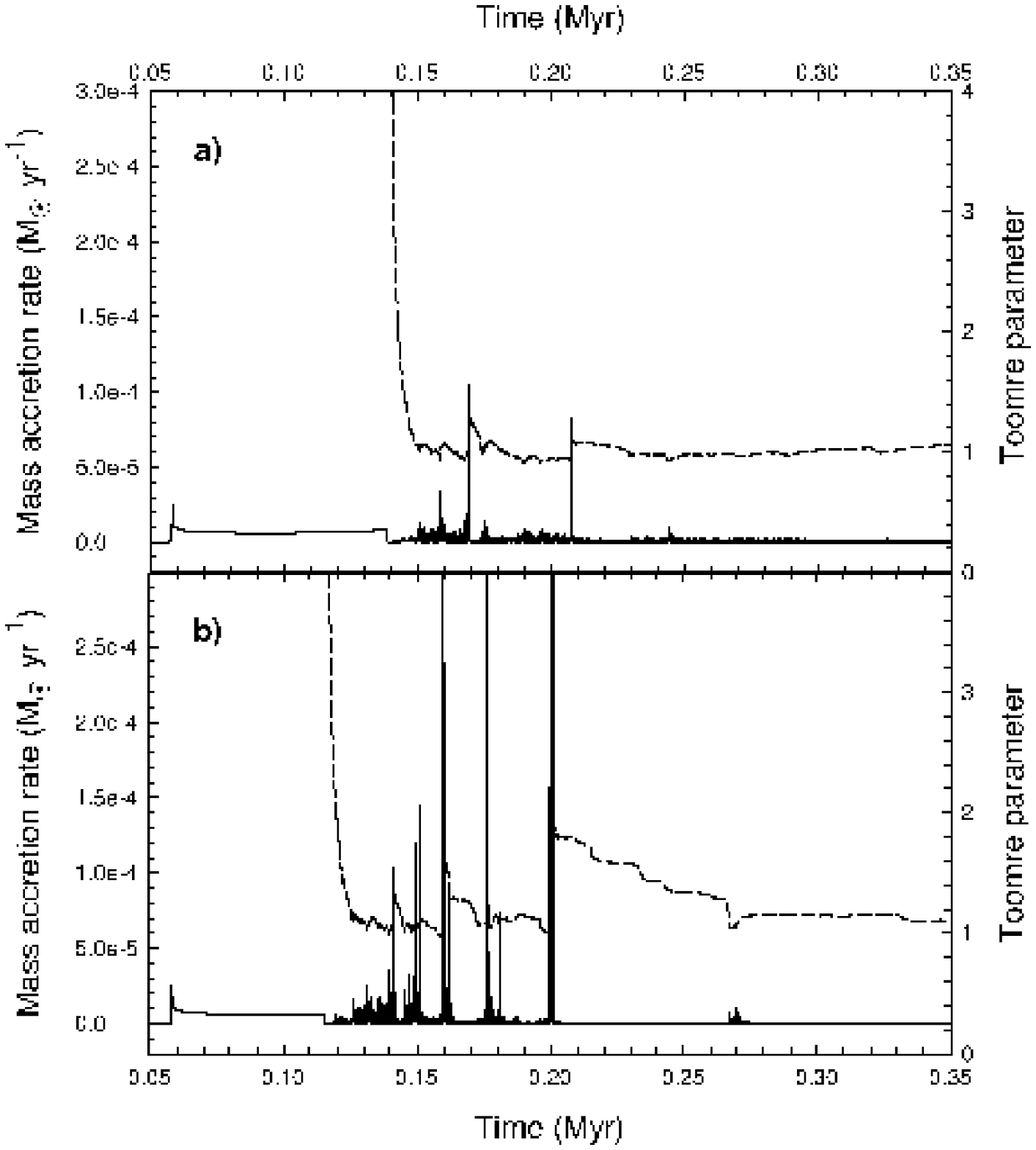}
      \caption{Temporal evolution of the mass accretion rate ({\em solid
      line}) and $Q$-parameter ({\em dashed line}). ($a$) Model~8, which has the same
      parameters as model~2 but a higher value of $\gamma=5/3$. ($b$)
      Model~9,
      which is identical to model~8 but has a higher initial rotational
      energy, i.e., $\beta=0.4\%$ }
         \label{fig13}
\end{figure}

The polytropic density-temperature relation~(\ref{tempr}) with $\gamma=5/3$
and $\Sigma_{\rm cr}=36.2$~g~cm$^{-2}$ ($n_{\rm cr}=10^{11}$~cm$^{-3}$)
is shown in Figure~\ref{fig11} by the dotted line. 
It is a reasonable fit to the results of MI (thick solid line) in the density range 
$n = 10^{11} {\rm cm}^{-3} - 10^{13} {\rm cm}^{-3}$ but yields
much greater temperatures at higher densities.
This can result in an additional heating of the protostellar disk 
in the optically thick regime and may suppress
the formation of dense protostellar/protoplanetary embryos. 
The solid line in Figure~\ref{fig13}a shows the evolution of the mass accretion rate
in model~8, which has the same parameters as model~2
but $\gamma=5/3$. It is evident that
an increased $\gamma$ considerably reduces the burst activity -- only two moderate-amplitude
bursts with $\dot{M}\la10^{-4}~M_\odot$~yr$^{-1}$ are now seen, in comparison to sixteen in model~2.

The suppressive effect of an elevated $\gamma$ (and by implication greater
disk temperature) on the embryo formation 
can be compensated to some extent by a greater infall rate onto the protostellar disk.
We demonstrate this by introducing a new model~9,
which has a higher initial rotational
energy ($\beta=0.4\%$) but is otherwise identical to model~8 
($\beta=0.275\%$). Given an increased initial rotational energy,
the protostellar disk forms at an earlier time and when the envelope mass
is greater. The solid line in Figure~\ref{fig13}b shows the mass accretion
rate onto the protostar $\dot{M}$ obtained in the $\beta=0.4\%$ model~9. It is
clear that the burst activity is greatly enhanced over that of the
$\beta=0.275\%$ model~8.
The difference in the envelope mass at the time of the disk formation 
in the two models is substantial: the $\beta=0.4\%$ model~9 has 
approximately $63\%$ of its total
mass in the envelope while the $\beta=0.275\%$ model~8 has only $46\%$.
Consequently, model~9 can sustain a high infall rate onto the protostellar
disk for a longer time than model~8, which favors the formation of dense
clumps and associated burst activity.
 
\section{Discussion}
\label{discussion}

Our parameter survey reveals that the basic mechanism of burst formation is 
robust and occurs under a variety of circumstances. Gravitational instability 
in the disk is driven by continuing infall from the envelope, 
which leads to spiral arms which fragment to form dense clumps.
Our long-term
integration reveals that the fate of these clumps
is to be driven onto the protostar, due to gravitational torques from their
interaction with the disk, especially the spiral arms. 

It is possible that disk heating sources (external, protostellar, or 
internal, e.g., shocks or compression) coupled with less efficient 
cooling may suppress the formation of the clumps. 
However, we find that the basic mechanism occurs even 
in models with either a low value of $\Sigma_{\rm cr}$ or with $\gamma = 5/3$, 
both of  which 
overestimate the gas temperature relative to that obtained in spherically
symmetric radiative transfer calculations (see Fig. \ref{fig11}). The strength and
frequency of the bursts is certainly dependent on the disk thermodynamics, but
also depends crucially on the envelope mass at the time of disk formation, since 
this process is ultimately driven by the mass accretion from the envelope.
We believe, on theoretical grounds, that the 
burst mode will occur as a part of the collapse of most (if not all) cloud cores, 
consistent with the observational
source statistics of FU Ori objects, which imply that every young
star undergoes some 10-20 bursts during its lifetime \citep{Hartmann2}.

Why has the burst mode of accretion not been discovered in any 
numerical simulations of disk evolution prior to our Paper I? 
We believe that there are two reasons:
(1) our model 
calculates the formation of a disk self-consistently from the collapse
of a cloud core,
and its continuing evolution as influenced by infall from the 
remnant core envelope;
(2) we are able to carry out a long-term time integration
(at least a few $\times 10^5$ yr after disk formation), in order to 
settle the fate of the clumps after they are formed.
Both of these advances are facilitated by our use of the 
two-dimensional thin-disk approximation, rather than three-dimensional
simulations which are currently far more resource-limited.

Several other groups have recently developed global 
three-dimensional numerical models
to study the onset of gravitational instability in
{\it isolated} and marginally unstable disks. These simulations
aim to understand whether
gas giant planets can be formed by direct gravitational
instability. Using either finite-difference  
\citep{Boss2,Pickett,Mejia}
or smoothed-particle hydrodynamics 
\citep{Rice,Mayer}
techniques, and either isothermal 
evolution or various recipes for heating, radiative cooling, and 
simplified radiative transfer, these simulations
often find that spiral structure gives birth to dense clumps
as in our model. Clump formation is reduced in models with prescribed
heating and cooling, and is even suppressed in some models
\citep{Mejia,Cai}.
All of the above models truncate the disk at an outer radius in the
range of 20 AU to 40 AU, with no account of outside influence.
Our view is that clump formation will likely occur
in any of these situations if the significant infall from the envelope 
during the early disk evolution is included. 
However, another important question is the 
survivability of the clumps, if they do form.
The three-dimensional simulations described above cannot follow the evolution
for more than $\sim 10^3$ yr, and therefore cannot settle this issue.
Our time-integration follows the disk evolution for $\ga 10^5$
yr, and each clump is followed for 
many orbit times, until its fate is settled. Our simulations show that
clumps are driven
onto the protostar (or are dispersed in a minority of cases), so that
there are no clumps remaining in the disk by the time that 99\% of the
envelope matter has been accreted. Furthermore, the gravitational 
instability has essentially ceased to operate by this late time.
This casts doubt on the possibility of gas giant planet formation
by direct gravitational instability.  
However, three-dimensional simulations and models which allow
the protostar to move off-center are also
necessary to confirm these results.
We believe that further progress in this field will come from
three-dimensional simulations of {\it nonisolated} disks
which have a high resolution, span a large range of length scales, 
and also integrate to very late times (at least $10^5$ yr after
disk formation). 
Increasingly sophisticated treatment of cooling,
heating, and radiative transfer, as some groups have begun to do,
are also essential for further progress.

The burst mode may hold the key to explaining the 
``luminosity problem'' of class I protostars in Taurus 
described by \cite{Kenyon2}; that is, the luminosities 
are much lower than predicted by the mass accretion rate
$0.975 \, \cs^3/G$ of the standard \cite{Shu77} model.
In an earlier paper \citep{VB1}, we explained the 
position of class 0 and class I protostars in a diagram of
envelope mass $M_{\rm env}$ versus bolometric luminosity
$L_{\rm bol}$ using evolutionary tracks from our spherical
collapse models. These models had the early, intermediate, and terminal
accretion phases, but no burst mode. The evolutionary tracks were in 
qualitative agreement
with the position of class I objects since they provided an explanation
for an apparent evolutionary decrease in $L_{\rm bol}$, due to the terminal
accretion phase. (In contrast, steady accretion at a rate $\propto \cs^3/G$
implies an indefinite increase in $L_{\rm bol}$.) However, we also
noted in \cite{VB1} that the model luminosities for Taurus class I sources
were still significantly greater than observed (the problem was
less severe for Ophiuchus class I sources). We believe that the
further decrease in $L_{\rm bol}$ implied by the entry of the accretion
into the quiescent phase of the burst mode may explain the remaining
luminosity problem. Both the effect of a core boundary (which 
causes the terminal phase of accretion) and rotation (which introduces
the burst mode) are likely needed to explain the luminosity 
problem. In other words, protostellar accretion needs to be understood
conceptually as an interplay of the three {\it phases} of accretion
(early, intermediate, and terminal - all of which exist in the absence of
rotation) with the two {\it modes} of 
accretion (smooth mode and burst mode - the latter is introduced 
due to rotation). 
A detailed analysis of evolutionary tracks is left for a separate study. 

The presence of a magnetic field is {\it not necessary} for the burst
mode to occur at all, as the results in \S\ \ref{rotate} demonstrate. 
The magnetic field in a supercritical core can moderate the 
burst activity but not suppress it. In all models of 
collapse, whether it is induced by gravity, turbulence, or ambipolar 
diffusion, the magnetic field is expected to be below the critical
value in order for collapse to ensue.
Our initial state is based on the
analytic best-fits to numerical models of supercritical
core collapse \citep{Basu}, and reflect the
approximate flux-freezing and a spatially uniform mass-to-flux ratio
in the early stages.  
Both assumptions are expected to break down
in the near-environment of a protostar, where significant 
magnetic flux redistribution will take place, due
to a combination of ambipolar diffusion \citep{CK98,Tassis}, 
ohmic dissipation \citep{Li}, and Hall diffusion \citep{Wardle}. 
A weakened magnetic field is likely to
enhance the strength of the bursts, as a comparison of our
model~2 and model~6 demonstrates. However, the magnetic flux-loss
from the inner region can introduce other, more complex, effects.
The outward movement of magnetic flux into regions where
it is better coupled to the neutrals can  
introduce a magnetic ``wall'' which retards accretion for a time
\citep{Li,CK98,Tassis}. This wall leads to a time-varying (fluctuating
by a factor $\sim 10$ above and below the mean value)
accretion rate onto the protostar in the calculations of \cite{Tassis},
who do not account for rotation or nonaxisymmetric effects. 
Future calculations will settle the extent to which nonaxisymmetric 
motions and magnetic interchange instability \citep{Stehle} can moderate such 
magnetically-induced accretion rate fluctuations.
Inclusion of dynamics in the third dimension (perpendicular to the disk plane)
is also critical to understand the full effect of magnetic fields,
since there is expected to be significant magnetic braking 
\citep{Krasnopolsky},
collapse-driven outflow \citep{Tomisaka2}, and magnetorotational instability
\citep{Fromang}, all of which can play a role in the angular momentum evolution
of the disk. The interaction of the stellar magnetic field with the 
disk gas and magnetic field is also important for gas
accretion \citep{Konigl,Shu94}. All in all, the 
interplay of disk formation and 
evolution (as calculated in this paper) with the various magnetic 
effects described above can be clarified through a series of future
multidimensional non-ideal MHD simulations.

\section{Summary}
\label{summary}
We have used the thin-disk approximation to 
calculate the runaway collapse and protostellar accretion phase
for an ensemble of cloud models, accounting for rotation, frozen-in
magnetic fields, and various relations for the density-temperature dependence.
We have elucidated the nature of the burst mode of accretion, first 
presented by \cite{VB3}. We find the following major results:

\begin{enumerate}

\item The collapse of flattened nonrotating cloud cores yield protostars whose 
mass accretion rate $\dot{M}$ is characterized by
three {\it phases}: an early phase of declining $\dot{M}$, an intermediate
phase of near-constant $\dot{M}$ (this phase does not occur if the
core is sufficiently compact), and a terminal phase of declining $\dot{M}$ due to 
a finite mass reservoir. These results agree qualitatively with earlier
results for the collapse of a spherical cloud. 

\item  The collapse of a rotating cloud core and self-consistent
formation of a protostellar disk leads to mass accretion in two 
distinct {\it modes}: a smooth mode, which is like the accretion in the
nonrotating model, followed by a burst mode which occurs upon the formation of
a protostellar disk. 
It is characterized by prolonged periods of low accretion rate from the
disk (the quiescent phases) that are
punctuated by intense bursts of accretion during which most of the
protostellar mass is accumulated.
The accretion bursts are associated with the formation of dense
protostellar/protoplanetary
embryos, which are later driven onto the protostar by
the gravitational torques that develop in the disk.
Gravitational instability in the disk that is driven by continuing infall 
from the cloud core envelope is responsible for the 
recurrent embryo formation.

\item The burst phenomenon is quite sensitive to the amount of 
angular momentum in the cloud, and the bursts are more frequent and intense
for greater values of $\beta$, the ratio of the magnitudes of rotational
and gravitational energy. For sufficiently low $\beta$, the disk may form
during the late part of the terminal accretion phase of the smooth mode,
and the infall onto the disk is not sufficient to create bursts.

\item The effect of a frozen-in magnetic field that is weaker than gravity
(i.e., the core is magnetically supercritical) is to moderate the burst
activity but not to suppress it. Magnetized disks have an intrinsically 
lower value of the critical $Q$ parameter for gravitational instability,
but are eventually driven unstable by 
infall from the envelope.

\item The effect of enhanced temperatures at high density, in accordance with 
(or actually exceeding) the results of detailed radiative transfer calculations
for spherical clouds, is to moderate the burst activity but not to suppress it.

\item Even in cases where the burst activity is nearly suppressed, 
we find that it can be made vigorous by moderate increases in the cloud size
or rotation rate. We conclude that the burst mode is a robust phenomenon 
which is likely to occur during the evolution of most (if not all) protostars.

\end{enumerate}

\begin{acknowledgements}
We thank S. Inutsuka for providing data on the 
density-temperature relation from his simulations.
This research was supported by the Natural Sciences and Engineering
Research Council of Canada. EIV gratefully acknowledges support
from a CITA National Fellowship.
\end{acknowledgements}

\appendix

\section{Numerical Tests}

A first test is of the 
van Leer advection scheme. We test it on a ``relaxation'' problem,
in which a thin disk of constant surface density is given a velocity 
field proportional 
to $r$ ($v_{r}=\nu_{0} r$), and the surface density is allowed to decrease.
For this problem, the analytic solution to the continuity equation is 
$\Sigma(t)=\Sigma_0
e^{-2 \nu_0 t}$ and it yields $\Sigma=6.14 \times 10^{-6}
\Sigma_0$ at $\nu_0 t=6$, i.e., when the surface density has
decreased by nearly six orders of magnitude. Our numerical code 
yields $\Sigma = 5.98\times 10^{-6} \Sigma_0$ at this time for a 
resolution of 256 logarithmically spaced radial grid points, 
so that the relative error is only 2.6\%.

\begin{figure}
\plotone{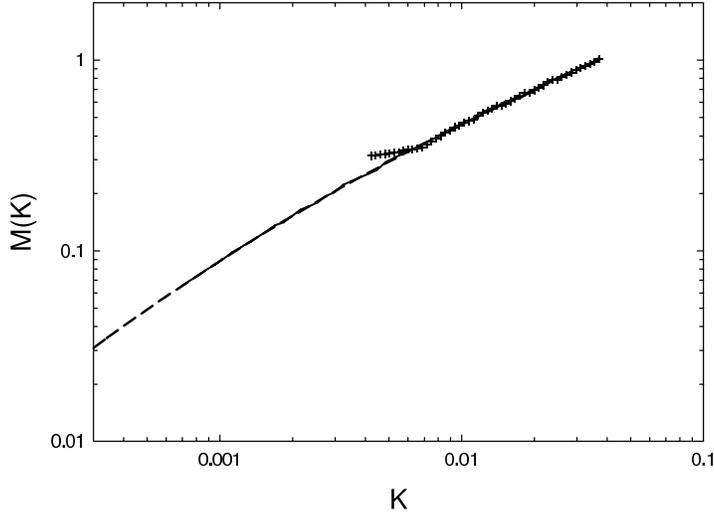}
      \caption{The specific angular momentum spectrum of a collapsing cloud
      core. The quantity $M(K)$ is the total mass in the cloud with specific
      angular momentum less than or equal to $K$. The dashed
      line shows the analytically derived spectrum of the initial state. The solid line
      shows $M(K)$ in our simulation after the formation of the protostar 
      but before the formation of the protostellar disk.
      The crosses show the spectrum after the formation of the
      protostellar disk, and reveal the significant 
      redistribution of angular momentum by nonaxisymmetric
      gravitational instabilities in the disk.}
         \label{fig14}
\end{figure}

An important concern for numerical studies of gravitational
collapse is the ability of a code to conserve specific angular momentum. 
A comprehensive test problem of specific angular momentum conservation 
that covers advection as well as pressure
and gravity terms in the momentum equations was designed by 
\citet{Norman}. For a fluid with no
mechanism for redistributing angular momentum, the mass $M(K)$ in the cloud
core with specific angular momentum ($\Omega r^2$) less than or equal to $K$
is a constant of the motion. A deviation from the initial spectrum $M(K)$
reveals a redistribution of angular momentum. For a uniformly rotating cloud 
($\Omega={\rm constant}$) with the radial surface density distribution given by 
equation~(\ref{density}), we have
\begin{equation}
M(K)=2 \pi \Sigma_0 r_0^2 \left( \sqrt{1+ {K\over K_0}} -1 \right),
\end{equation} 
where $K_0=\Omega r_0^2$. 
We consider a uniformly rotating ($\Omega=1.0$~km~s$^{-1}$~pc$^{-1}$) cloud core
with otherwise the same parameters as model~2. 
The initial theoretical spectrum $M(K)$ is shown in Figure~\ref{fig14}
by the dashed line. 
The solid line plots $M(K)$ computed at $t=0.07$~Myr, which is 
$\approx 5000$~yr after the formation
of the central protostar. At this time, the protostellar disk has not yet
formed and the flow of gas onto the protostar proceeds 
in an axisymmetric manner.
It is clear that $M(K)$ merges with the initial spectrum. This indicates virtually
no angular momentum redistribution due to either physical or numerical reasons.
We note that a part of the cloud core with low angular momentum has been
accreted into the sink cell.
When the protostellar disk forms at $t=0.107$~Myr, the 
gravitational torques associated with spiral arms 
produce a considerable deviation in the spectrum $M(K)$. This deviation
is apparent in Figure~\ref{fig14}, where we plot with crosses the spectrum 
at $t=0.115$~Myr.

\clearpage 

\begin{table}
\caption{Parameters of model cloud cores\tablenotemark{a} \label{table1}}
\vspace{3 pt}
\begin{tabular}{llllll}
\tableline\tableline
Model & $r_{\rm out}$ & $\Omega_0$ & $\alpha$ & $\gamma$ & $\Sigma_{\rm
cr}$ \\
\tableline
 1 & $10^4$ & 0    & 0 & 1.4 & 36.2\\
 2 & $10^4$ & 1.5  & 0 & 1.4 & 36.2 \\
 3 & $10^4$ & 1.35 & 0 & 1.4 & 36.2\\
 4 & $10^4$ & 1.2  & 0 & 1.4  & 36.2\\
 5 & $1.4\times10^4$ & 1.2  & 0 & 1.4 & 36.2\\
 6 & $10^4$ & 0    & 0.3 & 1.4 & 36.2\\
 7 & $10^4$ & 1.5    & 0 & 1.4  & 11.6\\
 8 & $10^4$ & 1.5    & 0 & 5/3 & 36.2\\
 9 & $10^4$ & 1.8    & 0 & 5/3 & 36.2\\
\tableline
\end{tabular}
\tablenotetext{a}{All distances are in AU, angular
velocities in km~s$^{-1}$~pc$^{-1}$, and surface densities in g~cm$^{-2}$.}
\end{table}


\begin{thebibliography}{}

\bibitem[Bacmann et al.(2000)]{Bacmann}
Bacmann, A., Andr\'{e}, P., Puget, J. L., et al. 2000, A\&A, 314, 625

\bibitem[Basu(1997)]{Basu}
Basu, S. 1997, ApJ, 485, 240 

\bibitem[Basu \& Mouschovias(1994)]{Basu94}
Basu, S., \& Mouschovias, T. Ch. 1994, ApJ, 432, 720

\bibitem[Basu \& Mouschovias(1995a)]{Basu95a}
\ul. 1995a, ApJ, 452, 386

\bibitem[Basu \& Mouschovias(1995b)]{Basu95b}
\ul. 1995b, ApJ, 453, 271

\bibitem[Bate et al.(2003)Bate, Bonnell, \& Bromm]{Bate}
Bate, M. R., Bonnell, I. A., \& Bromm, V. 2003, MNRAS, 339, 577

\bibitem[Bell \& Lin(1994)]{Bell94}
Bell, K. R., \& Lin, D. N. C. 1994, \apj, 427, 987

\bibitem[Bell et al.(1995)]{Bell95}
Bell, K. R., Lin, D. N. C., Hartmann, L. W., \& Kenyon, S. J. 1995, 
\apj, 444, 376

\bibitem[Binney \& Tremaine(1987)]{BT}
Binney, J., \& Tremaine, S. 1987, Galactic Dynamics (Princeton:
Princeton Univ. Press)

\bibitem[Bonnell \& Bastien(1992)]{Bonnell}
Bonnell, I., \& Bastien, P. 1992, ApJ, 401, L31 

\bibitem[Boss(1998)]{Boss} 
Boss, A. P. 1998, ApJ, 503, 923

\bibitem[Boss(2003)]{Boss2}
\ul. 2003, ApJ, 599, 577  

\bibitem[Cai et al.(2006)]{Cai}
Cai, K., Durisen, R. H., Michael, S., Boley, A. C., Mej\'ia, A. C.,
Pickett, M. K., \& D'Alessio, P. 2006, \apjl, 636, L149


\bibitem[Ciolek \& K\"onigl(1998)]{CK98}
Ciolek, G. E., \& K\"onigl, A. 1998, \apj, 504, 257 

\bibitem[Ciolek \& Mouschovias(1993)]{Ciolek} 
Ciolek, G. E., \& Mouschovias, T. Ch. 1993, \apj, 454, 194 

\bibitem[Clarke et al.(1990)]{Clarke}
Clarke, C. J., Lin, D. N. C., \& Pringle, J. E. 1990, MNRAS, 242, 439 

\bibitem[Fatuzzo et al.(2004)Fatuzzo, Adams, \& Myers]{Fatuzzo}
Fatuzzo, M., Adams, F. C., \& Myers, P. C. 2004, \apj, 615, 813

\bibitem[Fiedler \& Mouschovias(1993)]{Fiedler} 
Fiedler, R. A., \& Mouschovias, T. Ch. 1993, \apj, 415, 680

\bibitem[Foster \& Chevalier(1993)]{FC}
Foster, P. N., \& Chevalier, R. A. 1993, \apj, 416, 303

\bibitem[Fromang et al.(2005)]{Fromang}
Fromang, S., Balbus, S. A., Terquem, C., \& De Villiers, J.-P. 2005,
\apj, 616, 364

\bibitem[Fukagawa et al.(2004)]{Fukagawa}
Fukagawa, M., Hayashi, M., Tamura, M. et al. 2004, ApJ, 605, L53

\bibitem[Grady et al.(2001)]{Grady}
Grady, C. A., Polomski, E. F., Henning, Th., et al. 2001, AJ, 122, 3396

\bibitem[Hartmann(1998)]{Hartmann}
Hartmann, L. 1998, Accretion Processes in Star Formation
(Cambridge: Cambridge Univ. Press)

\bibitem[Hartmann \& Kenyon(1996)]{Hartmann2}
Hartmann, L., \& Kenyon, S. 1996, \araa, 34, 207

\bibitem[Henriksen et al.(1997)Henriksen, Andr\'{e}, \& Bontemps]{Henriksen}
Henriksen, R., Andr\'{e}, P., \& Bontemps, S. 1997, A\&A, 323, 549

\bibitem[Herbig(1977)]{Herbig} 
Herbig, G. H. 1977, ApJ, 217, 693

\bibitem[Hunter(1977)]{Hunter}
Hunter, C. 1977, \apj, 218, 834

\bibitem[Kenyon \& Hartmann(1995)]{Kenyon2}
Kenyon, S. J., \& Hartmann, L. 1995, ApJS, 101, 117

\bibitem[Kenyon et al.(1990)]{Kenyon}
Kenyon, S. J., Hartmann, L. W., Strom, K. M., \& Strom, S. E. 1990, AJ,
99, 869

\bibitem[K{\"o}nigl(1991)]{Konigl}
K{\"o}nigl, A. 1991, \apjl, 370, L39

\bibitem[Krasnopolsky \& K{\"o}nigl(2002)]{Krasnopolsky}
Krasnopolsky, R., \& K{\"o}nigl, A. 2002, ApJ, 580, 987

\bibitem[Larson(1984)]{Larson84}
Larson, R. B. 1984, \mnras, 206, 197

\bibitem[Larson(2003)]{Larson}
\ul. 2003, Rep. Prog. Phys. 66, 1651

\bibitem[Laughlin \& Bodenheimer(1994)]{Laughlin}
Laughlin, G., \& Bodenheimer, P. 1994, ApJ, 436, 335 

\bibitem[Li \& McKee(1996)]{Li}
Li, Z.-Y., \& McKee, C. F. 1996, \apj, 464, 373

\bibitem[Li \& Shu(1997)]{LS}
Li, Z.-Y., \& Shu, F. H. 1997, ApJ, 475, 237

\bibitem[Lin \& Papaloizou(1985)]{Lin}
Lin, D. N. C., \& Papaloizou, J. C. B. 1985, in Protostars and Planets~II,
ed. D. C. Black  \& M. C. Matthews (Tucson: Univ. Arizona Press), 981 

\bibitem[Masunaga \& Inutsuka(2000)]{Masunaga}
Masunaga, H., \& Inutsuka, S. 2000, ApJ, 531, 350

\bibitem[Masunaga et al.(1998)Masunaga, Miyama, \& Inutsuka]{Masunaga98}
Masunaga, H., Miyama, S. M., \& Inutsuka, S. 1998, ApJ, 495, 346

\bibitem[Mayer et al.(2004)]{Mayer}
Mayer, L., Quinn, T., Wadsley, J., \& Stadel, J. 2004, \apj, 609, 1045

\bibitem[Mej\'ia et al.(2005)]{Mejia}
Mej\'ia, A. C., Durisen, R. H., Pickett, M. K., \& Cai, K. 2005, ApJ, 619, 1098

\bibitem[Nakamura \& Hanawa(1997)]{Nakamura}
Nakamura, F., \& Hanawa, T. 1997, \apj, 480, 701

\bibitem[Nakano \& Nakamura(1978)]{Nakano}
Nakano, T., \& Nakamura, T. 1978, PASJ, 30, 671

\bibitem[Narita et al.(1984)Narita, Hayashi, \& Miyama]{Narita}
Narita,, S. Hayashi, C., \& Miyama, S. M. 1984, Prog. Theor. Phys., 72, 1118

\bibitem[Nelson et al.(1998)]{Nelson}
Nelson, A. F., Benz, W., Adams, F. C., \& Arnett, D. 1998, ApJ, 502, 342

\bibitem[Norman et al.(1980)Norman, Wilson, \& Barton]{Norman}
Norman, M. L., Wilson, J. R., \& Barton, R. T. 1980, ApJ, 239, 968

\bibitem[Pickett et al.(2003)]{Pickett}
Pickett, B. K., Mej\'ia, A. C., Durisen, R. H., Cassen, P. M.,
Berry, D., \& Link, R. P. 2003, \apj, 590, 1060

\bibitem[Polyachenko et al.(1997)Polyachenko, Polyachenko, \& Strel'nikov]{Polyachenko}
Polyachenko, V. L., Polyachenko, E. V., \& Strel'nikov, A. V. 1997, Astron.
Zhurnal, 23, 551 (translated Astron. Lett. 23, 483)

\bibitem[Rice et al.(2003)]{Rice}
Rice, W. K. M., Armitage, P. J., Bate, M. R., \& Bonnell, I. A. 2003,
\mnras, 339, 1025

\bibitem[Shakura \& Sunyaev(1973)]{SS}
Shakura, N. I., \& Sunyaev, R. A. 1973, A\&A, 24, 337

\bibitem[Shu(1977)]{Shu77}
Shu, F. H. 1977, ApJ, 214, 488

\bibitem[Shu \& Li(1997)]{SL}
Shu, F. H., \& Li, Z.-Y. 1997, ApJ, 475, 251

\bibitem[Shu et al.(1994)]{Shu94}
Shu, F. H., Najita, J., Ostriker, E., Wilkin, F., Ruden, S., \& Lizano, S.
1994, \apj, 429, 781

\bibitem[Stehle \& Spruit(2002)]{Stehle}
Stehle, R., \& Spruit, H. C. 2002, \mnras, 323, 587

\bibitem[Stone \& Norman(1992)]{SN}
Stone, J. M., \& Norman, M. L. 1992, ApJS, 80, 753 

\bibitem[Tassis \& Mouschovias(2005)]{Tassis}
Tassis, K., \& Mouschovias, T. Ch. 2005, \apj, 618, 783

\bibitem[Tomisaka(1996)]{Tomisaka}
Tomisaka, K. 1996, PASJ, 48, L97

\bibitem[Tomisaka(2002)]{Tomisaka2}
\ul. 2002, \apj, 575, 306

\bibitem[Tomley et al.(1994)Tomley, Steinman-Cameron, \& Cassen]{Tomley}
Tomley, L., Steiman-Cameron, T. Y., \& Cassen, P. 1994, ApJ, 422, 850 

\bibitem[van Leer(1977)]{vanLeer} 
van Leer, B. 1977, JCP, 23, 276

\bibitem[Vorobyov \& Basu(2005a)]{VB1}
Vorobyov, E. I., \&  Basu, S. 2005a, MNRAS, 360, 675

\bibitem[Vorobyov \& Basu(2005b)]{VB2}
\ul. 2005b, MNRAS, 363, 1361

\bibitem[Vorobyov \& Basu(2005c)]{VB3}
\ul. 2005c, ApJ, 633, L137 (Paper I)

\bibitem[Wardle(1999)]{Wardle}
Wardle, M. 1999, \mnras, 307, 849

\bibitem[Ward-Thompson et al.(1999)Ward-Thompson, Motte, \& Andr\'{e}]{WT}
Ward-Thompson, D., Motte, F., \& Andr\'{e}, P. 1999, MNRAS, 305, 143

\bibitem[Whitworth \& Ward-Thompson(2001)]{WW01}
Whitworth, A. P., \& Ward-Thompson, D. 2001, ApJ, 547, 317


\end{thebibliography}
\end{document}